\documentclass[12pt,onecolumn]{IEEEtran}
\usepackage{amssymb}
\usepackage{amsfonts}
\usepackage{amsmath}
\usepackage{algorithm}
\usepackage{algorithmic}
\usepackage{tikz}
\usepackage{graphicx,subfigure,cite,multicol,multirow,diagbox,booktabs,array}

\ifCLASSINFOpdf
\else
\fi

\begin{document}
%
\title{ Enhanced Secrecy Rate Maximization for  Directional Modulation Networks via IRS\\ }

\author{Feng~Shu,~Jiayu~Li,~Mengxing~Huang,~Weiping~Shi,~Yin~Teng,~Jun~Li,\\~Yongpeng~Wu,~and~Jiangzhou~Wang~\emph{Fellow},~\emph{IEEE}

\thanks{This work was supported in part by the National Natural Science Foundation of China (Nos. 61771244) \emph{(Corresponding authors: Mengxing Huang~,~Weiping~Shi,~and Feng Shu,)}.}
\thanks{Feng~Shu  and Mengxing Huang are  with the School of Information and Communication Engineering, Hainan University,~Haikou,~570228, China.}
\thanks{Feng~Shu,~Jiayu~Li,~Weiping~Shi,~Yin~Teng,~and Jun Li are with the School of Electronic and Optical Engineering, Nanjing University of Science and Technology, 210094, China.}
\thanks{Y. Wu is with the Shanghai Key Laboratory of Navigation and LocationBased Services, Shanghai Jiao Tong University, Minhang 200240, China (e-mail:,yongpeng.wu2016@gmail.com).}
\thanks{Jiangzhou Wang is with the School of Engineering and Digital Arts, University of Kent, Canterbury CT2 7NT, U.K. Email: (e-mail: j.z.wang@kent.ac.uk).}

}
\maketitle

\begin{abstract}
Intelligent reflecting surface (IRS) is of low-cost and energy-efficiency and will be a promising technology for the future wireless communications like sixth generation. To address the problem of conventional directional modulation (DM) that Alice only transmits single confidential bit stream (CBS) to Bob with multiple antennas in a line-of-sight channel, IRS is proposed to create friendly multipaths for DM such that two CBSs can be transmitted from Alice to Bob. This will significantly enhance the secrecy rate (SR) of DM. To maximize the SR (Max-SR), a general non-convex optimization problem is formulated with the unit-modulus constraint of IRS phase-shift matrix (PSM), and the general alternating iterative (GAI)  algorithm is proposed to jointly obtain the transmit beamforming vectors (TBVs) and PSM by  alternately optimizing one and fixing another. To reduce its high complexity, a low-complexity iterative algorithm for Max-SR is proposed by placing the constraint of null-space (NS) on the TBVs, called NS projection (NSP). Here, each CBS is transmitted separately in the NSs of other CBS and AN channels. Simulation results show that the SRs of the proposed GAI and NSP can approximately double that of IRS-based DM with single CBS for massive IRS in the high signal-to-noise ratio region.
\end{abstract}

\maketitle

\section{Introduction}
\IEEEPARstart{W}{ith} the commercialization of the fifth-generation (5G) and the requirements of sixth-generation (6G) pre-research, physical layer security increasingly becomes an extremely important and prominent problem.
Techniques such as massive multiple-input multiple-output ({MIMO}), millimeter wave (mmWave) mobile communication and hybrid beamforming have been investigated in cellular systems, internet of things ({IoT}), unmanned aerial vehicle ({UAV}), and satellite communications\cite{massivemimo, TS-MMWAVE, hbsurvey}.
However, the network energy consumption and hardware cost still remain critical issues. For example, 5G system has a much higher energy consumption  than 4G system\cite{WANG-OFDMA1,WANG-OFDMA2}. Therefore, the importance of green communication becomes increasingly significant for the future wireless communications.  Many related technologies are in the pace of research, such as simultaneous wireless information and power transfer ({SWIPT}), which can enhance the energy efficiency and solve energy-limited issues of wireless networks\cite{liqxax,WuQ-viewGreen5G,MMGreen5G}.

For physical layer security, \cite{Wyner1975} proposed the concept of secrecy capacity in a discrete memoryless wiretap channel.
With the aid of artificial noise (AN), the security can be improved against the overhearing of potential eavesdroppers\cite{AN-wiretap-YSH}.
As one of the most attractive technology in physical layer security, directional modulation (DM) is to apply signal processing methods like beamforming and {AN} in radio frequency (RF)  frontend or baseband, so that the signal in the desired direction can be restored as completely as possible, while the constellation diagram of signal in the undesired direction is distorted \cite{DM-PA}. Traditional DM synthesis formed an orthogonal vector or projection matrix in the null space (NS) of channel along the desired direction, which can be seen as a kind of NS projection (NSP) schemes\cite{YD-avector}. \cite{sun2020energy-efficient} proposed an energy-efficient alternating iterative scheme and discussed the secure energy efficiency for DM system.\cite{LJY-DM} has considered the secure performance analysis related to the quantization error caused by phase shifters, which inspires the hardware cost in the practical application of DM.  In \cite{HU2016-RDM, WU2016-RDMBC, MU-MIMO-ZHU, DM-XU}, the authors proposed robust DM synthesis schemes in several different scenarios as single-desired user, multi-user (MU) broadcasting, MU-MIMO and multicast in the presence of  direction of arrival (DOA) measurement errors.
To achieve the high-resolution estimation of direction of arrival (DOA) for practical DM, \cite{DOA-QIN} proposed three high-performance estimators of DOA for hybrid MIMO structure. Futhermore, a practical DM scheme with random frequency diverse array was proposed in \cite{RFDA-HJS}, inspiring a new concept birth of secure and precise wireless transmission to achieve  a higher-level physical layer security \cite{Wu-SPTDM}.

As wireless networks develop rapidly, a large number of active devices will result in a serious problem of energy consumption. Therefore, how to introduce passive devices and achieve a trade-off between spectrum utilization and energy efficiency with low hardware costs becomes a necessity for achieving sustainable wireless network evolution. Moreover, the improvement of propagation environment and coverage of base station (BS) also become one of the important research areas of next-generation wireless communications. Its main aim is to create a smart environment for transmitting BS signals. Intelligent reflecting surface (IRS) has now become a promising and emerging technology with great potential of significant energy consumption reduction and spectrum efficient enhancement\cite{wuqq-zongshu}. It is a planer array consisting of a large number of reconfigurable passive elements, where each of them can be controlled by an attached smart controller and thus induce a certain phase shift independently on incident signal to change the reflected signal propagation. This reveals the potential of enhancing the signal transmission and coverage.
Due to the passive forwarding and full-duplex characteristics without self-interference, IRS can play an important role in coverage improvement, spectrum and energy efficiency enhancement, and the complexity and power consumption reduction of wireless networks.

Existing algorithms for IRS-based system implementation focus on the improvement of energy efficiency and secure capacity. The phase-shifters of IRS with constant modulus makes it difficult to solve the optimization problem. In \cite{huangc-2018GC}, the authors proposed the energy efficiency maximization of IRS-aided multi-input single-output ({MISO}) system when the  phase-shifters of IRS are of low resolution, while \cite{huangc-2019} investigated the case of infinite resolution.
The authors in \cite{wuqq1} and \cite{wuqq3-discrete} focused on the design of transmit beamforming by active antenna array and reflect beamforming by passive IRS to minimize the total transmit power, and discussed the cases of continuous and discrete phase-shifter. The efficient algorithms with semifinite relaxation (SDR) and alternating optimization techniques in \cite{wuqq1} were proposed to make a tradeoff between the system performance and computational complexity. As for the IRS-aided MIMO system, \cite{IRS-ZY-MIMO} aimed to characterize the fundamental capacity limit and developed efficient alternating optimization algorithms both in narrow band  and broadband scenarios. In \cite{IRS-mmMIMO}, IRS was  proposed to  be employed in  mmWave massive MIMO in practice.
Since all the above works focused on one-way communications, \cite{IRS-ZY-MIMOtw} proposed the sum rate optimization of IRS-aided full-duplex MIMO two-way communications through jointly optimizing the source precoders and the IRS phase-shift matrix.

Apart from the above traditional communication situations, IRS can also be applied in some special cases, such as cognitive radio systems, {UAV} communications and {SWIPT}\cite{IRS-CR-XG,IRS-CR-LZ,IRS-SL-UAV,SHI2019SWIPT,IRS-MIMO-SWIPT}. Some literature has made a  special investigation of the impact of the number of  element of IRS on communication performance, as \cite{IRS-O-number} analyzed the minimum limit of IRS element number to achieve a certain transmission rate.
Moreover, the path-loss impact related to IRS was discussed in \cite{IRS-pathloss-DN} and \cite{IRS-O-pathloss}. \cite{IRS-pathloss-DN} established the path loss model and analyzed the performance through experiments, while \cite{IRS-O-pathloss} applied physical optics technology to analyze the path loss expressions related to IRS link in the far field. The above research makes IRS more feasible in practice.

In IRS-based secure wireless communications, confidential message (CM) can be transmitted by direct path and reflected by reflect path. However, the CM could be leaked to the undesired directions, which may reduce the secure performance. In this case, the scheme of IRS-based in secure communication should be treated seriously. \cite{IRS-CM-secure} investigated an IRS-based secure system with multi-antenna transmitter Alice, single-antenna receiver Bob and single-antenna eavesdropper Eve. The authors applied alternating optimization and SDR methods to maximize the secrecy rate  (SR). \cite{SH-SRMIRS} proposed an iterative algorithm for designing the transmit covariance matrix in a closed form and IRS phase-shift matrix in a semi-closed form, respectively.
As for IRS-based MIMO secure communication, the authors in \cite{IRS-LD-secureMIMO} and \cite{IRS-LD-ANsecureMIMO} studied the SR maximization in the case of the direct link between transmitter, receiver and eavesdropper.
\cite{IRS-AN,IRS-JSAC-ROBUSTAN,PAN_SEC_MIMO,IRS-LD-ANsecureMIMO} investigated the potential of AN in IRS-aided communications in which AN can be an effective means to help improve the SR with IRS deployed in practice, especially for multi-eavesdroppers.

In traditional {DM} networks, the signal should be transmitted in a line-of-sight (LOS) channel to enhance the directivity of transmission. This will lead to a drawback of DM that only single bit stream may be sent from Alice to Bob.
To overcome the limitation, employing IRS in DM network will generate multipath  to achieve a smart environment of transmitting controllable multiple parallel bit streams from Alice to Bob. In other words, due to IRS, spatial multiplexing gains are created for DM.  This means that the SR performance can be dramatically improved.  Moreover, IRS can ensure a low energy consumption of DM system compared with other active forwarding devices like relays, which will make a good balance between spectrum efficiency and energy efficiency.
Compared with traditional IRS-based {MIMO} secure communication in \cite{IRS-LD-secureMIMO}, AN in DM system not only interferes with eavesdropping, but also remains the problem of interference to legitimate users through the reflective path. Therefore, it is necessary to design a reasonably secure transmission scheme for the IRS-based DM MIMO network.

In this paper, we consider an IRS-based {DM} network, where  all Alice, Bob and Eve are employed with multiple antennas. In a direct way and a reflective way with the help of IRS,   the suitably phase-shifted versions of transmitted signals are  forwarded towards Bob and interfere with Eve seriously. Additionally, IRS is equipped with a large number of controllable reflecting elements with continuous phase-shifters. The contributions of this paper are summarized as follows:

\begin{enumerate}
 \item To overcome the limitation of DM that Alice only transmit single confidential bit stream (CBS) to Bob with multiple antennas in  LOS channel, the IRS-based DM network is proposed. With the help of IRS, useful multipaths are created between Alice and Bob. As such, multiple parallel  CBSs may be transmitted from Alice to Bob. This will result in a significant improvement in SR. As shown in what follows,  if two parallel independent CBSs are sent from Alice to Bob, the proposed IRS-based DM framework can harvest up to 75\% SR gain over single CBS as the number of IRS elements tends to large-scale.

 \item To maximize the SR (Max-SR) of system, a general algorithm is proposed. Since the objective problem is non-convex for the unit-modulus constraint of IRS phase-shift matrix, we propose the general alternating iterative (GAI)  algorithm to jointly obtain the transmit beamforming vectors and IRS phase-shift matrix by optimizing one and fixing another. It is assumed that AN is in the NS of Alice-to-Bob channel and Alice-to-IRS channel, that is, only interferes with Eve. In the proposed GAI,  the closed-form expression of transmit beamforming vector corresponding to each CBS  is derived, and the iterative gradient ascent algorithm is adopted  to optimize the IRS phase-shift matrix. The proposed GAI performs much better than random phase, no-IRS, and IRS with single CBS in terms of SR. Its SR approximately doubles that of the IRS with single CBS.

 \item To reduce the high computational complexity of the proposed GAI,   a low-complexity iterative Max-SR is proposed by imposing NS constraints on all beamforming vectors. Below, this method is short for NSP. Here, each CBS is transmitted separately in the NSP of other CBS channels transmitter-to-receiver links. It is interesting that the IRS phase-shift matrix has  a semi-closed form. In the risk of a little SR performance, this method can achieve a low computation complexity, especially when the number of IRS elements is high. Compared to the proposed GAI, the proposed NSP shows a little SR performance loss but its low-complexity is very attractive. Moreover, by simulation,  we find the location of IRS has an important impact on the SR performance of methods and is preferred to  be close to  Alice or Bob in order to enhance better security.
    %
\end{enumerate}

The remainder of this paper is organized as follows. Section \ref{S2} describes the system model and secrecy maximization problem. In Section \ref{S3}, the general alternating iterative algorithm is proposed. Section \ref{S4} describes another low-complexity algorithm for special scenario. Simulation results and related analysis are presented in Section \ref{S5}. Finally, we make our conclusions in Section \ref{S6}.

\emph{Notations:} throughout the paper, matrices, vectors, and scalars are denoted by letters of bold upper case, bold lower case, and lower case, respectively. Signs $(\cdot)^T$, $(\cdot)^\ast$, $(\cdot)^H$, $(\cdot)^{-1}$, $(\cdot)^{\dagger}$ and $|\cdot|$ denote transpose, conjugate, conjugate transpose, inverse, pseudo-inverse and matrix determinant, respectively. $\textbf{I}_N$ denotes the $N\times N$ identity matrix, $\mathbf{0}_{N\times M}$ denotes the $N\times M$ matrix of all zeros.

\section{System Model and Problem Formulation}\label{S2}
\subsection{System Model Description}
\begin{figure}[htb]
  \centering
  \includegraphics[width=0.5\textwidth]{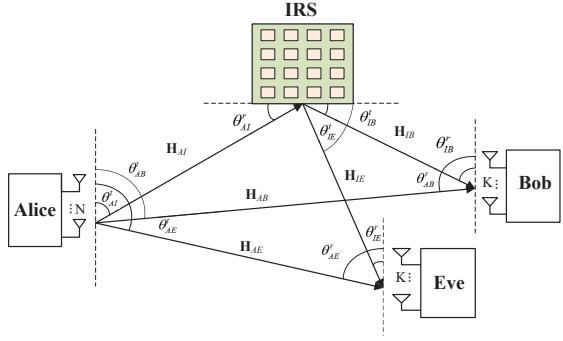}
  \caption{Block diagram for IRS-based DM network.}
  \label{systemmodel}
\end{figure}
As shown in Fig.~\ref{systemmodel} , we consider a system, where Alice is equipped with $N$ antennas, IRS is equipped with $M$ low-cost passive reflecting elements, Bob and Eve are equipped with $K$ antennas, respectively. In the following , we assume that  the IRS reflects signal only one time. In this paper, we assume there exists the LOS path. The transmit baseband signal is expressed as
\begin{equation}\label{s}
\mathbf{s}=\sqrt{\beta_1 P_s}\mathbf{v}_{1}x_{1}+\sqrt{\beta_2 P_s}\mathbf{v}_{2}x_{2}+\sqrt{(1-\beta_1-\beta_2)P_s}\mathbf{P}_{AN}\mathbf{z},
\end{equation}
where $P_s$ is the total transmit power, $\beta_1$, $\beta_2$ and $(1-\beta_1-\beta_2)$ are the power allocation parameters of {CMs} and {AN}, respectively. $\mathbf{v}_1\in \mathbb{C}^{N\times 1}$ and $\mathbf{v}_2\in \mathbb{C}^{N\times 1}$ are the beamforming vector of forcing the two CMs to the desired user Bob, where $\mathbf{v}_{1}^{H}\mathbf{v}_{1}=1$, $\mathbf{v}_{2}^{H}\mathbf{v}_{2}=1$. Beamforming vectors $\mathbf{v}_{AN}\in \mathbb{C}^{N\times 1}$ are the beamforming vectors of leading AN to the undesired direction, where $\mathbf{v}_{AN}^{H}\mathbf{v}_{AN}=1$. $x_1$ and $x_2$ are CM which satisfy $\mathbb{E}\left[\|x_1\|^2\right]=1$, $\mathbb{E}\left[\|x_2\|^2\right]=1$, and $\mathbf{z}$ is vector AN with complex Gaussian distribution, i.e., $\mathbf{z}\sim\mathcal{C}\mathcal{N}(0,~\mathbf{I}_{N})$.
The received signal at Bob is given by
\begin{align}\label{yb}
\mathbf{y}_B&=\left(\sqrt{g_{AIB}}\mathbf{H}^{H}_{IB}\boldsymbol{\Theta}\mathbf{H}_{AI}+\sqrt{g_{AB}}\mathbf{H}^{H}_{AB} \right)\mathbf{s}+\mathbf{n}_{B} \\
&=\sqrt{\beta_1 P_s}\left(\sqrt{g_{AIB}}\mathbf{H}^{H}_{IB}\boldsymbol{\Theta}\mathbf{H}_{AI}+\sqrt{g_{AB}}\mathbf{H}^{H}_{AB} \right)\mathbf{v}_{1}x_{1} \nonumber\\
&+\sqrt{\beta_2 P_s}\left(\sqrt{g_{AIB}}\mathbf{H}^{H}_{IB}\boldsymbol{\Theta}\mathbf{H}_{AI}+\sqrt{g_{AB}}\mathbf{H}^{H}_{AB} \right)\mathbf{v}_{2}x_{2} \nonumber\\
&+\sqrt{(1-\beta_1-\beta_2) P_s}\left(\sqrt{g_{AIB}}\mathbf{H}^{H}_{IB}\boldsymbol{\Theta}\mathbf{H}_{AI}+\sqrt{g_{AB}}\mathbf{H}^{H}_{AB} \right)\mathbf{P}_{AN}\mathbf{z}+\mathbf{n}_{B},\nonumber
\end{align}
where $\mathbf{H}_{IB}\in\mathbb{C}^{M\times K}$ represents the IRS-to-Bob channel, $\boldsymbol{\Theta}=\text{diag}(e^{j\phi_1},\cdots,e^{j\phi_m},\cdots, e^{j\phi_M})$ is a diagonal matrix with the phase shift $\phi_m$ incurred by the $m$-th reflecting element of the IRS,  $\mathbf{H}_{AI}\in\mathbb{C}^{M\times N}$ represents  the Alice-to-IRS channel, $\mathbf{H}_{AB}\in\mathbb{C}^{N\times K}$ represents Alice-to-Bob channel, and $\mathbf{n}_B\sim\mathcal{C}\mathcal{N}(\mathbf{0},\sigma_B^2\mathbf{I}_{K})$ denotes the complex additive white Gaussian noise (AWGN) at Bob.
 $g_{AB}$ denotes the path loss coefficient between Alice and Bob, whereas $g_{AIB}$ is the equivalent path loss coefficient of Alice-to-IRS channel and IRS-to-Bob channel. Similarly, the received signal at Eve can be written as
\begin{align}\label{ye}
\mathbf{y}_E&=\left(\sqrt{g_{AIE}}\mathbf{H}^{H}_{IE}\boldsymbol{\Theta}\mathbf{H}_{AI}+\sqrt{g_{AE}}\mathbf{H}^{H}_{AE} \right) \mathbf{s}+\mathbf{n}_{E}\\
&=\sqrt{\beta_1 P_s}\left(\sqrt{g_{AIE}}\mathbf{H}^{H}_{IE}\boldsymbol{\Theta}\mathbf{H}_{AI}+\sqrt{g_{AE}}\mathbf{H}^{H}_{AE} \right)\mathbf{v}_{1}x_{1} \nonumber\\
&+\sqrt{\beta_2 P_s}\left(\sqrt{g_{AIE}}\mathbf{H}^{H}_{IE}\boldsymbol{\Theta}\mathbf{H}_{AI}+\sqrt{g_{AE}}\mathbf{H}^{H}_{AE} \right)\mathbf{v}_{2}x_{2} \nonumber\\
&+\sqrt{(1-\beta_1-\beta_2) P_s}\left(\sqrt{g_{AIE}}\mathbf{H}^{H}_{IE}\boldsymbol{\Theta}\mathbf{H}_{AI}+\sqrt{g_{AE}}\mathbf{H}^{H}_{AE}\right)\mathbf{P}_{AN}\mathbf{z}+\mathbf{n}_{E},\nonumber
\end{align}
where $\mathbf{H}_{IE}\in\mathbb{C}^{M\times K}$ represents the IRS-to-Eve channel, $\mathbf{H}_{AE}\in\mathbb{C}^{N\times K}$ represents the Alice-to-Eve channel, and $\mathbf{n}_E\sim\mathcal{C}\mathcal{N}(\mathbf{0},\sigma_E^2\mathbf{I}_{K})$ denotes AWGN at Bob. Here, $g_{AIE}$ and $g_{AE}$ denote the path loss coefficient between Alice and Eve, where $g_{AIE}$ is the equivalent path loss coefficient of Alice-to-IRS channel and IRS-to-Eve channel, $g_{AE}$ is the path loss coefficient of Alice-to-Eve channel. In the following, we assume that $\sigma_B^2=\sigma_E^2=\sigma^2$.

Assuming that AN  is only transmitted to Eve for interference, then $\mathbf{P}_{AN}$ should satisfy the condition that
\begin{align}\label{panzf}
\mathbf{H}_{AI}\mathbf{P}_{AN}=\mathbf{0}_{M\times N},~\mathbf{H}_{AB}^{H}\mathbf{P}_{AN}=\mathbf{0}_{K\times N}.
\end{align}
Let us define a large virtual CM channel as follows
\begin{align}\label{H-CM}
\mathbf{H}_{CM} = \left[ \mathbf{H}^T_{AI}~
\mathbf{H}_{AB}^\ast
\right]^T,
\end{align}
then $\mathbf{P}_{AN}$ can be expressed as
\begin{align}\label{PAN}
\mathbf{P}_{AN}=\mathbf{I}_{N}-\mathbf{H}_{CM}^{H}\left[\mathbf{H}_{CM}\mathbf{H}_{CM}^{H}\right]^{\dagger}\mathbf{H}_{CM}.
\end{align}
In this case, (\ref{yb}) and (\ref{ye}) can be rewritten by applying (\ref{PAN}) as,
\begin{align}\label{yban}
\mathbf{y}_B&=\sqrt{\beta_1 P_s}\left(\sqrt{g_{AIB}}\mathbf{H}^{H}_{IB}\boldsymbol{\Theta}\mathbf{H}_{AI}+\sqrt{g_{AB}}\mathbf{H}^{H}_{AB} \right)\mathbf{v}_{1}x_{1} \\
&+\sqrt{\beta_2 P_s}\left(\sqrt{g_{AIB}}\mathbf{H}^{H}_{IB}\boldsymbol{\Theta}\mathbf{H}_{AI}+\sqrt{g_{AB}}\mathbf{H}^{H}_{AB} \right)\mathbf{v}_{2}x_{2}+\mathbf{n}_{B},\nonumber
\end{align}
\begin{align}\label{yean}
\mathbf{y}_E&=\sqrt{\beta_1 P_s}\left(\sqrt{g_{AIE}}\mathbf{H}^{H}_{IE}\boldsymbol{\Theta}\mathbf{H}_{AI}+\sqrt{g_{AE}}\mathbf{H}^{H}_{AE} \right)\mathbf{v}_{1}x_{1} \\
&+\sqrt{\beta_2 P_s}\left(\sqrt{g_{AIE}}\mathbf{H}^{H}_{IE}\boldsymbol{\Theta}\mathbf{H}_{AI}+\sqrt{g_{AE}}\mathbf{H}^{H}_{AE} \right)\mathbf{v}_{2}x_{2} \nonumber\\
&+\sqrt{(1-\beta_1-\beta_2) P_s}\sqrt{g_{AE}}\mathbf{H}^{H}_{AE}\mathbf{P}_{AN}\mathbf{z}+\mathbf{n}_{E}.\nonumber
\end{align}

\subsection{Secrecy Rate Maximization Problem}
We jointly optimize beamforming vectors and IRS phase-shift matrix $\boldsymbol{\Theta}$ based on the secrecy rate maximization scheme.The achievable rates from Alice to Bob and to Eve can be expressed as
\begin{align}\label{RBNO}
R_B&=\log_2\left|\mathbf{I}_K+\frac{1}{\sigma^2}\big(\beta_1 P_{s}\mathbf{H}_{B}\mathbf{v}_{1}\mathbf{v}_{1}^{H}\mathbf{H}_{B}^{H}
+\beta_{2}P_{s}\mathbf{H}_{B}\mathbf{v}_{2}\mathbf{v}_{2}^{H}\mathbf{H}_{B}^{H}\big)\right|\\
&=\log_2\left|\mathbf{I}_K+\mathbf{H}_{B1}\mathbf{v}_{1}\mathbf{v}_{1}^{H}\mathbf{H}_{B1}^{H}+\mathbf{H}_{B2}\mathbf{v}_{2}\mathbf{v}_{2}^{H}\mathbf{H}_{B2}^{H}\right|\nonumber
\end{align}
and
\begin{align}\label{RENO}
R_E&=\log_2\left|\mathbf{I}_K+\frac{\beta_1 P_{s}\mathbf{H}_{E}\mathbf{v}_{1}\mathbf{v}_{1}^{H}\mathbf{H}_{E}^{H}
+\beta_{2}P_{s}\mathbf{H}_{E}\mathbf{v}_{2}\mathbf{v}_{2}^{H}\mathbf{H}_{E}^{H}}
{(1-\beta_1-\beta_2)P_sg_{AE}\mathbf{H}_{AE}^{H}\mathbf{P}_{AN}\mathbf{P}_{AN}^{H}\mathbf{H}_{AE}+\sigma^2\mathbf{I}_{K}}\right|\\
&=\log_2\left|\mathbf{I}_K+\left(\mathbf{H}_{E1}\mathbf{v}_{1}\mathbf{v}_{1}^{H}\mathbf{H}_{E1}^{H}+\mathbf{H}_{E2}\mathbf{v}_{2}\mathbf{v}_{2}^{H}\mathbf{H}_{E2}^{H}\right)\mathbf{B}^{-1}\right|,\nonumber
\end{align}
where $\mathbf{H}_{B}= \sqrt{g_{AIB}}\mathbf{H}^{H}_{IB}\Theta\mathbf{H}_{AI}+\sqrt{g_{AB}}\mathbf{H}^{H}_{AB}$, $\mathbf{H}_{E}= \sqrt{g_{AIE}}\mathbf{H}^{H}_{IE}\Theta\mathbf{H}_{AI}+\sqrt{g_{AE}}\mathbf{H}^{H}_{AE}$.
The achievable SR $R_{s}$ can be written as
\begin{align}\label{RSNO}
R_{s}&=\max \left\{0, R_B-R_E \right\}\\
&=\log_2\left|\frac{\mathbf{I}_K+\mathbf{H}_{B1}\mathbf{v}_{1}\mathbf{v}_{1}^{H}\mathbf{H}_{B1}^{H}+\mathbf{H}_{B2}\mathbf{v}_{2}\mathbf{v}_{2}^{H}\mathbf{H}_{B2}^{H}}
{\mathbf{I}_K+\left(\mathbf{H}_{E1}\mathbf{v}_{1}\mathbf{v}_{1}^{H}\mathbf{H}_{E1}^{H}+\mathbf{H}_{E2}\mathbf{v}_{2}\mathbf{v}_{2}^{H}\mathbf{H}_{E2}^{H}\right)\mathbf{B}^{-1}}\right|.\nonumber
\end{align}
The achievable SR given by optimization problem can be formulated as follows:

\begin{subequations}\label{P0}
\begin{align}
\mathrm{(P0):}&\max_{\mathbf{v}_1,\mathbf{v}_2,\boldsymbol{\Theta}}~~R_s(\mathbf{v}_1,\mathbf{v}_2,\boldsymbol{\Theta})\label{P00}\\
&\text{s.t.}~~\mathbf{v}_1^H\mathbf{v}_1=1,\mathbf{v}_2^H\mathbf{v}_2=1,\label{P01}\\
&~~~~~|\Theta_{i}|=1,\arg(\Theta_{i})\in[0,2\pi),~i=1,\cdots, M,\label{P02}
\end{align}
\end{subequations}
where $\Theta_{i}$ is the $i$-th diagonal of $\boldsymbol{\Theta}$. It is hard to solve the problem since the unit modulus constraint is hard to handle. In this case, we propose the alternating algorithm to calculate the beamforming vectors and IRS phase shift matrix separatively.

\section{Proposed high-performance GAI-based Max-SR method}\label{S3}

In this section, we propose an optimal alternating algorithm for secrecy rate maximization problem to determine the beamforming vectors for CM and AN, and IRS phase-shift matrix $\boldsymbol{\Theta}$.
To simplify the expression of $R_s$, let us first define
\begin{align}
&\mathbf{H}_{B1}=\frac{\sqrt{\beta_1P_s}}{\sigma}\big(\sqrt{g_{AIB}}\mathbf{H}^{H}_{IB}\boldsymbol{\Theta}\mathbf{H}_{AI}+\sqrt{g_{AB}}\mathbf{H}^{H}_{AB} \big),\\
&\mathbf{H}_{B2}=\frac{\sqrt{\beta_2P_s}}{\sigma}\big(\sqrt{g_{AIB}}\mathbf{H}^{H}_{IB}\boldsymbol{\Theta}\mathbf{H}_{AI}+\sqrt{g_{AB}}\mathbf{H}^{H}_{AB} \big),\\
&\mathbf{C}_{B1}=\mathbf{H}_{B1}\mathbf{v}_1\mathbf{v}_1^{H}\mathbf{H}_{B1}^{H},\\
&\mathbf{C}_{B2}=\mathbf{H}_{B2}\mathbf{v}_2\mathbf{v}_2^{H}\mathbf{H}_{B2}^{H},
\end{align}
for  Bob and
\begin{align}
&\mathbf{H}_{E1}=\frac{\sqrt{\beta_1P_s}}{\sigma}\big(\sqrt{g_{AIE}}\mathbf{H}^{H}_{IE}\boldsymbol{\Theta}\mathbf{H}_{AI}+\sqrt{g_{AE}}\mathbf{H}^{H}_{AE} \big),\\
&\mathbf{H}_{E2}=\frac{\sqrt{\beta_2P_s}}{\sigma}\big(\sqrt{g_{AIE}}\mathbf{H}^{H}_{IE}\boldsymbol{\Theta}\mathbf{H}_{AI}+\sqrt{g_{AE}}\mathbf{H}^{H}_{AE} \big),\\
&\mathbf{C}_{E1}=\mathbf{H}_{E1}\mathbf{v}_1\mathbf{v}_1^{H}\mathbf{H}_{E1}^{H},\\
&\mathbf{C}_{E2}=\mathbf{H}_{E2}\mathbf{v}_2\mathbf{v}_2^{H}\mathbf{H}_{E2}^{H},\\
&\mathbf{B}=\frac{(1-\beta_1-\beta_2)P_s g_{AE}}{\sigma^2}\mathbf{H}_{AE}^{H}\mathbf{P}_{AN}\mathbf{P}_{AN}^{H}\mathbf{H}_{AE}+\mathbf{I}_K,\label{B}
\end{align}
for  Eve.
\subsection{Optimize the beamforming vectors $\mathbf{v}_1$ and $\mathbf{v}_2$  given the IRS phase-shift matrix $\boldsymbol{\Theta}$}
To simplify the expression of $R_s$ related to beamforming vectors, we regard $\boldsymbol{\Theta}$ as a given constant matrix, and define that
\begin{align}
{R}_{B}(\mathbf{v}_1)&\overset{(a)}{=}\log_2|\mathbf{I}_{K}+\mathbf{C}_{B2}|+\label{RBv1}\log_2|\mathbf{I}_{K}+(\mathbf{I}_{K}+\mathbf{C}_{B2})^{-1}\mathbf{H}_{B1}\mathbf{v}_1\mathbf{v}_1^{H}\mathbf{H}_{B1}^{H}|\\
&\overset{(b)}{=}
\log_2|\mathbf{I}_{K}+\mathbf{C}_{B2}|+\log_2\big(1+\mathbf{v}_1^{H}\mathbf{H}_{B1}^{H}(\mathbf{I}_{K}+\mathbf{C}_{B2})^{-1}\mathbf{H}_{B1}\mathbf{v}_1\big),\nonumber\\
{R}_{E}(\mathbf{v}_1)&\overset{(a)}{=}\log_2|\mathbf{I}_{K}+\mathbf{C}_{E2}\mathbf{B}^{-1}|+\label{REv1}\log_2|\mathbf{I}_{K}+(\mathbf{I}_{K}+\mathbf{C}_{E2}\mathbf{B}^{-1})^{-1}\mathbf{H}_{E1}\mathbf{v}_1\mathbf{v}_1^{H}\mathbf{H}_{E1}^{H}\mathbf{B}^{-1}|\\
&\overset{(b)}{=}\log_2|\mathbf{I}_{K}+\mathbf{C}_{E2}\mathbf{B}^{-1}|+\log_2\big(1+\mathbf{v}_1^{H}\mathbf{H}_{E1}^{H}\mathbf{B}^{-1}(\mathbf{I}_{K}+\mathbf{C}_{E2}\mathbf{B}^{-1})^{-1}\mathbf{H}_{E1}\mathbf{v}_1\big),\nonumber
\end{align}
where $(a)$ holds due to the fact that $|\mathbf{X}\mathbf{Y}|=|\mathbf{X}||\mathbf{Y}|$ and $(b)$ holds due to $|\mathbf{I}_M+\mathbf{X}\mathbf{Y}|=|\mathbf{I}_N+\mathbf{Y}\mathbf{X}|$ for $\mathbf{X}\in \mathbb{C}^{M\times N}$ and $\mathbf{Y}\in \mathbb{C}^{N\times M}$.
Rewrite (\ref{RSNO}) by applying (\ref{RBv1}) and (\ref{REv1}),
\begin{align}\label{RSEV}
R_s(\mathbf{v}_1)&=\log_2|\mathbf{I}_{K}+\mathbf{C}_{B2}|-\log_2|{\mathbf{I}_{K}+\mathbf{C}_{E2}\mathbf{B}^{-1}}|+\log_2\frac{\mathbf{v}_1^{H}\tilde{\mathbf{C}}_{B2}\mathbf{v}_1}{\mathbf{v}_1^{H}\tilde{\mathbf{C}}_{E2}\mathbf{v}_1},
\end{align}
where $\tilde{\mathbf{C}}_{B2}=\mathbf{I}_N+\mathbf{H}_{B1}^{H}(\mathbf{I}_{K}+\mathbf{C}_{B2})^{-1}\mathbf{H}_{B1}$,
$\tilde{\mathbf{C}}_{E2}=\mathbf{I}_N+\mathbf{H}_{E1}^{H}\mathbf{B}^{-1}(\mathbf{I}_{K}+\mathbf{C}_{E2}\mathbf{B}^{-1})^{-1}\mathbf{H}_{E1}$.
Since the first two items of (\ref{RSEV}) are independent of $\mathbf{v}_1$, the subproblem to optimize $\mathbf{v}_1$ can be expressed as follows:
\begin{align}\label{RSV1}
\mathrm{(P0-1):}\max_{\mathbf{v}_1}~~\frac{\mathbf{v}_1^{H}\tilde{\mathbf{C}}_{B2}\mathbf{v}_1}{\mathbf{v}_1^{H}\tilde{\mathbf{C}}_{E2}\mathbf{v}_1}~~~~~~~~\text{s.t.}~~\mathbf{v}_1^{H}\mathbf{v}_1=1.
\end{align}
According to the Rayleigh-Ritz theorem, the optimal $\mathbf{v}_1$ can be obtained from the eigenvector corresponding to the largest eigenvalue of the matrix $\tilde{\mathbf{C}}_{E2}^{-1}\tilde{\mathbf{C}}_{B2}$.

Similarly, given the determined or known $\mathbf{v}_1$ and ${\boldsymbol{\Theta}}$, let us define $\tilde{\mathbf{C}}_{B1}=\mathbf{I}_N+\mathbf{H}_{B2}^{H}(\mathbf{I}_{K}+\mathbf{C}_{B1})^{-1}\mathbf{H}_{B2}$ and
$\tilde{\mathbf{C}}_{E1}=\mathbf{I}_N+\mathbf{H}_{E2}^{H}\mathbf{B}^{-1}(\mathbf{I}_{K}+\mathbf{C}_{E1}\mathbf{B}^{-1})^{-1}\mathbf{H}_{E2}$. The subproblem to optimize $\mathbf{v}_2$ can be expressed as follows:
\begin{align}\label{RSV2}
\mathrm{(P0-2):}\max_{\mathbf{v}_2}~~\frac{\mathbf{v}_2^{H}\tilde{\mathbf{C}}_{B1}\mathbf{v}_2}{\mathbf{v}_2^{H}\tilde{\mathbf{C}}_{E1}\mathbf{v}_2}~~~~~~~~\text{s.t.}~~\mathbf{v}_2^{H}\mathbf{v}_2=1.
\end{align}
According to the Rayleigh-Ritz theorem, the optimal $\mathbf{v}_2$ can be obtained from the eigenvector corresponding to the largest eigenvalue of the matrix $\tilde{\mathbf{C}}_{E1}^{-1}\tilde{\mathbf{C}}_{B1}$.

\subsection{Optimize IRS phase-shift matrix $\boldsymbol{\Theta}$ given the beamforming vectors}
To simplify the expression of $R_s$ in this subsection, we define the IRS phase-shift vector  containing all the elements on the diagonal of $\boldsymbol{\Theta}$, that is,
\begin{align}\label{thetade}
&\boldsymbol{\theta}=[e^{j\phi_1}, \cdots, e^{j\phi_m}, \cdots, e^{j\phi_M}]^T,
\boldsymbol{\Theta}=\text{diag}\{\boldsymbol{\theta}\}.
\end{align}
Letting $\theta_i=e^{j\phi_i}$ be the $i$-th element of $\boldsymbol{\theta}$,  the IRS phase-shift vector $\boldsymbol{\theta}$ should satisfy the condition that
\begin{align}\label{thetacon}
|\theta_{i}|=1,\arg(\theta_{i})\in[0,2\pi),~i=1,\cdots, M.
\end{align}
Here, let us define
\begin{align}
&\mathbf{g}_1=\mathbf{H}_{AI}\mathbf{v}_1, \mathbf{g}_2=\mathbf{H}_{AI}\mathbf{v}_2,\label{g}\\
&\mathbf{h}_{B1}=\frac{\sqrt{\beta_1 P_s g_{AB}}}{\sigma}\mathbf{H}_{AB}^{H}\mathbf{v}_1,
\mathbf{h}_{B2}=\frac{\sqrt{\beta_2 P_s g_{AB}}}{\sigma}\mathbf{H}_{AB}^{H}\mathbf{v}_2,\label{hb}\\
&\mathbf{h}_{E1}=\frac{\sqrt{\beta_1 P_s g_{AE}}}{\sigma}\mathbf{H}_{AE}^{H}\mathbf{v}_1,
\mathbf{h}_{E2}=\frac{\sqrt{\beta_2 P_s g_{AE}}}{\sigma}\mathbf{H}_{AE}^{H}\mathbf{v}_2.\label{he}
\end{align}
Given that
\begin{align}\label{HB1V1}
\mathbf{H}_{B1}\mathbf{v}_1&=\frac{\sqrt{\beta_1 P_s}}{\sigma}(\sqrt{g_{AIB}}\mathbf{H}_{IB}^{H}\boldsymbol{\Theta}\mathbf{H}_{AI}\mathbf{v}_1+\sqrt{g_{AB}}\mathbf{H}_{AB}^{H}\mathbf{v}_1)\\
&\overset{(c)}{=}\frac{\sqrt{\beta_1 P_s g_{AIB}}}{\sigma}\mathbf{H}_{IB}^{H} \text{diag}\{\mathbf{g}_1\}\boldsymbol{\theta}+\mathbf{h}_{B1},\nonumber
\end{align}
where $(c)$ holds due to the fact that $\text{diag}\{\mathbf{a}\}\mathbf{b}=\text{diag}\{\mathbf{b}\}\mathbf{a}$ for $\mathbf{a}, \mathbf{b}\in\mathbb{C}^{M\times 1}$.
To simplify the above equation, we define
\begin{align}
\mathbf{T}_{B1}=\frac{1}{\sigma}\sqrt{\beta_1 P_s g_{AIB}}\mathbf{H}_{IB}^{H} \text{diag}\{\mathbf{g}_1\},\label{TB1}\\
\mathbf{T}_{B2}=\frac{1}{\sigma}\sqrt{\beta_2 P_s g_{AIB}}\mathbf{H}_{IB}^{H} \text{diag}\{\mathbf{g}_2\},\label{TB2}\\
\mathbf{T}_{E1}=\frac{1}{\sigma}\sqrt{\beta_1 P_s g_{AIE}}\mathbf{H}_{IE}^{H} \text{diag}\{\mathbf{g}_1\},\label{TE1}\\
\mathbf{T}_{E2}=\frac{1}{\sigma}\sqrt{\beta_2 P_s g_{AIE}}\mathbf{H}_{IE}^{H} \text{diag}\{\mathbf{g}_2\}.\label{TE2}
\end{align}
Then (\ref{HB1V1}) can be rewritten as
\begin{align}\label{thb1}
\mathbf{H}_{B1}\mathbf{v}_1=\mathbf{T}_{B1}\boldsymbol{\theta}+\mathbf{h}_{B1}.
\end{align}
For the sake of simplicity, we define $\mathbf{t}_{hb1}\triangleq\mathbf{H}_{B1}\mathbf{v}_1$. Similarly, the expression like $\mathbf{H}_{B1}\mathbf{v}_1$ can also be defined as
$\mathbf{t}_{hb2}\triangleq\mathbf{H}_{B2}\mathbf{v}_2$,
$\mathbf{t}_{he1}\triangleq\mathbf{H}_{E1}\mathbf{v}_1$,
$\mathbf{t}_{he2}\triangleq\mathbf{H}_{E2}\mathbf{v}_2$,
that is,
\begin{align}\label{th}
\mathbf{t}_{hb1}=\mathbf{T}_{B1}\boldsymbol{\theta}+\mathbf{h}_{B1}, \mathbf{t}_{hb2}=\mathbf{T}_{B2}\boldsymbol{\theta}+\mathbf{h}_{B2},\\
\mathbf{t}_{he1}=\mathbf{T}_{E1}\boldsymbol{\theta}+\mathbf{h}_{E1},
\mathbf{t}_{he2}=\mathbf{T}_{E2}\boldsymbol{\theta}+\mathbf{h}_{E2}.
\end{align}
In this case, we rewrite (\ref{RBNO}) and (\ref{RENO}) as
\begin{align}
{R}_{B}(\boldsymbol{\theta})
&=\log_2\left|\mathbf{I}_{K}+\mathbf{t}_{hb1}\mathbf{t}_{hb1}^{H}+\mathbf{t}_{hb2}\mathbf{t}_{hb2}^{H}\right|\label{RBt}\\\
&=\log_2\left|\left(\mathbf{I}_{K}+\mathbf{t}_{hb2}\mathbf{t}_{hb2}^{H}\right)\left(\mathbf{I}_{K}+\left(\mathbf{I}_{K}+\mathbf{t}_{hb2}\mathbf{t}_{hb2}^{H}\right)^{-1}\mathbf{t}_{hb1}\mathbf{t}_{hb1}^{H}\right)\right|\nonumber\\
&=\log_2\left(1+\mathbf{t}_{hb2}^{H}\mathbf{t}_{hb2}\right)+\log_2\left( 1+\mathbf{t}_{hb1}^{H}\left(\mathbf{I}_{K}+\mathbf{t}_{hb2}\mathbf{t}_{hb2}^{H}\right)^{-1}\mathbf{t}_{hb1}\right)\nonumber
\end{align}
and
\begin{align}
{R}_{E}(\boldsymbol{\theta})
&=\log_2\left|\mathbf{I}_{K}+(\mathbf{t}_{he1}\mathbf{t}_{he1}^{H}+\mathbf{t}_{he2}\mathbf{t}_{he2}^{H})\mathbf{B}^{-1}\right|\label{REt}\\
&=\log_2\left|\left(\mathbf{I}_{K}+\mathbf{t}_{he2}\mathbf{t}_{he2}^{H}\mathbf{B}^{-1}\right)\left(\mathbf{I}_{K}+\left(\mathbf{I}_{K}+\mathbf{t}_{he2}\mathbf{t}_{he2}^{H}\mathbf{B}^{-1}\right)^{-1}\mathbf{t}_{he1}\mathbf{t}_{he1}^{H}\mathbf{B}^{-1}\right)\right|\nonumber\\
&=\log_2\left(1+\mathbf{t}_{he2}^{H}\mathbf{B}^{-1}\mathbf{t}_{he2}\right)+\log_2\left( 1+\mathbf{t}_{he1}^{H}\mathbf{B}^{-1}\left(\mathbf{I}_{K}+\mathbf{t}_{he2}\mathbf{t}_{he2}^{H}\mathbf{B}^{-1}\right)^{-1}\mathbf{t}_{he1}\right).\nonumber
\end{align}

The SR in terms of $\boldsymbol{\theta}$ can be rewritten as
\begin{align}\label{RBET}
(\ref{RBt})-(\ref{REt})=\log_2\frac{f_1(\boldsymbol{\theta})f_2(\boldsymbol{\theta})}{g_1(\boldsymbol{\theta})g_2(\boldsymbol{\theta})},
\end{align}
where
\begin{align}
f_1(\boldsymbol{\theta})&=1+\mathbf{t}_{hb2}^{H}\mathbf{t}_{hb2},\\
f_2(\boldsymbol{\theta})&=1+\mathbf{t}_{hb1}^{H}\left(\mathbf{I}_{K}+\mathbf{t}_{hb2}\mathbf{t}_{hb2}^{H}\right)^{-1}\mathbf{t}_{hb1}\\
&\overset{(d)}{=}1+\mathbf{t}_{hb1}^{H}\left(\mathbf{I}_{K}-\mathbf{t}_{hb2}(1+\mathbf{t}_{hb2}^{H}\mathbf{t}_{hb2})^{-1}\mathbf{t}_{hb2}^{H}\right)\mathbf{t}_{hb1}\nonumber\\
&=1+\mathbf{t}_{hb1}^H\mathbf{t}_{hb1}-\frac{\mathbf{t}_{hb1}^H\mathbf{t}_{hb2}\mathbf{t}_{hb2}^H\mathbf{t}_{hb1}}{1+\mathbf{t}_{hb2}^H\mathbf{t}_{hb2}},\nonumber\\
g_1(\boldsymbol{\theta})&=1+\mathbf{t}_{he2}^{H}\mathbf{B}^{-1}\mathbf{t}_{he2},\\
g_2(\boldsymbol{\theta})&=1+\mathbf{t}_{he1}^{H}\mathbf{B}^{-1}\left(\mathbf{I}_{K}+\mathbf{t}_{he2}\mathbf{t}_{he2}^{H}\mathbf{B}^{-1}\right)^{-1}\mathbf{t}_{he1}\\
&\overset{(d)}{=}1+\mathbf{t}_{he1}^{H}\mathbf{B}^{-1}\left(\mathbf{I}_{K}-\frac{\mathbf{t}_{he2}\mathbf{t}_{he2}^{H}\mathbf{B}^{-1}}{1+\mathbf{t}_{he2}^{H}\mathbf{B}^{-1}\mathbf{t}_{he2}}\right)\mathbf{t}_{he1}\nonumber\\
&=1+\mathbf{t}_{he1}^{H}\mathbf{B}^{-1}\mathbf{t}_{he1}-\frac{\mathbf{t}_{he1}^{H}\mathbf{B}^{-1}\mathbf{t}_{he2}\mathbf{t}_{he2}^{H}\mathbf{B}^{-1}\mathbf{t}_{he1}}{1+\mathbf{t}_{he2}^{H}\mathbf{B}^{-1}\mathbf{t}_{he2}},\nonumber
\end{align}
where $(d)$ holds the fact that $(\mathbf{I}_{M}+\mathbf{X}\mathbf{Y})^{-1}=\mathbf{I}_M-\mathbf{X}(\mathbf{I}_{N}+\mathbf{Y}\mathbf{X})^{-1}\mathbf{Y}$ for $\mathbf{X}\in\mathbb{C}^{M\times N}$ and $\mathbf{Y}\in\mathbb{C}^{N\times M}$. To simplify the expression, let us define that
\begin{align}
&f(\boldsymbol{\theta})=f_1(\boldsymbol{\theta})f_2(\boldsymbol{\theta})=f_{t1}(\boldsymbol{\theta})-f_{t2}(\boldsymbol{\theta}),\\
&g(\boldsymbol{\theta})=g_1(\boldsymbol{\theta})g_2(\boldsymbol{\theta})=g_{t1}(\boldsymbol{\theta})-g_{t2}(\boldsymbol{\theta}),
\end{align}
where $f_{t1}(\boldsymbol{\theta})=(1+\mathbf{t}_{hb1}^H\mathbf{t}_{hb1})(1+\mathbf{t}_{hb2}^H\mathbf{t}_{hb2})$,
$f_{t2}(\boldsymbol{\theta})=\mathbf{t}_{hb1}^{H}\mathbf{t}_{hb2}\mathbf{t}_{hb2}^H\mathbf{t}_{hb1}$,
$g_{t1}(\boldsymbol{\theta})=(1+\mathbf{t}_{he1}^H\mathbf{B}^{-1}\mathbf{t}_{he1})(1+\mathbf{t}_{he2}^H\mathbf{B}^{-1}\mathbf{t}_{he2})$,
$g_{t2}(\boldsymbol{\theta})=\mathbf{t}_{he1}^{H}\mathbf{B}^{-1}\mathbf{t}_{he2}\mathbf{t}_{he2}^H\mathbf{B}^{-1}\mathbf{t}_{he1}$.
Then the subproblem to optimize $\boldsymbol{\theta}$ can be formulated as
\begin{align}\label{RST}
\mathrm{(P0-3):}\max_{\boldsymbol{\theta}}~~\frac{f(\boldsymbol{\theta})}{g(\boldsymbol{\theta})}=\frac{f_{t1}(\boldsymbol{\theta})-f_{t2}(\boldsymbol{\theta})}{g_{t1}(\boldsymbol{\theta})-g_{t2}(\boldsymbol{\theta})}~~~~~~~~~~\text{s.t.}~~(\ref{thetacon}).
\end{align}
Since (\ref{RBET}) is a non-convex function of $\boldsymbol{\theta}$, and all elements in $\boldsymbol{\theta}$ are of constant modulus constraint, thus, a gradient ascent (GA) method is used to compute the {IRS} phase-shift matrix $\boldsymbol{\Theta}=\text{diag}\{\boldsymbol{\theta}\}$. The gradient of the objective function in (\ref{RST}) with respect to $\boldsymbol{\theta}$ can be expressed as
\begin{align}\label{deltathetano}
\nabla_{\boldsymbol{\theta}}=\frac{f^{'}(\boldsymbol{\theta})g(\boldsymbol{\theta})-f(\boldsymbol{\theta})g^{'}(\boldsymbol{\theta})}{g^2(\boldsymbol{\theta})},
\end{align}
where
\begin{align}
f^{'}(\boldsymbol{\theta})&=f_{t1}^{'}(\boldsymbol{\theta})-f_{t2}^{'}(\boldsymbol{\theta}),~g^{'}(\boldsymbol{\theta})=g_{t1}^{'}(\boldsymbol{\theta})-g_{t2}^{'}(\boldsymbol{\theta}),\\
f_{t1}^{'}(\boldsymbol{\theta})&=(1+\mathbf{t}_{hb2}^H\mathbf{t}_{hb2})(\mathbf{T}_{B1}^{H}\mathbf{T}_{B1}\boldsymbol{\theta}+\mathbf{T}_{B1}^{H}\mathbf{h}_{B1})+(1+\mathbf{t}_{hb1}^H\mathbf{t}_{hb1})(\mathbf{T}_{B2}^{H}\mathbf{T}_{B2}\boldsymbol{\theta}+\mathbf{T}_{B2}^{H}\mathbf{h}_{B2}),\\
f_{t2}^{'}(\boldsymbol{\theta})&=\mathbf{t}_{hb2}^{H}\mathbf{t}_{hb1}(\mathbf{T}_{B2}^{H}\mathbf{T}_{B1}\boldsymbol{\theta}+\mathbf{T}_{B2}^{H}\mathbf{h}_{B1})+\mathbf{t}_{hb1}^{H}\mathbf{t}_{hb2}(\mathbf{T}_{B1}^{H}\mathbf{T}_{B2}\boldsymbol{\theta}+\mathbf{T}_{B1}^{H}\mathbf{h}_{B2}),\\
g_{t1}^{'}(\boldsymbol{\theta})&=(1+\mathbf{t}_{he2}^H\mathbf{B}^{-1}\mathbf{t}_{he2})(\mathbf{T}_{E1}^{H}\mathbf{B}^{-1}\mathbf{T}_{E1}\boldsymbol{\theta}+\mathbf{T}_{E1}^{H}\mathbf{B}^{-1}\mathbf{h}_{E1})\\
&+(1+\mathbf{t}_{he1}^H\mathbf{B}^{-1}\mathbf{t}_{he1})(\mathbf{T}_{E2}^{H}\mathbf{B}^{-1}\mathbf{T}_{E2}\boldsymbol{\theta}+\mathbf{T}_{E2}^{H}\mathbf{B}^{-1}\mathbf{h}_{E2}),\nonumber\\
g_{t2}^{'}(\boldsymbol{\theta})&=\mathbf{t}_{he2}^{H}\mathbf{B}^{-1}\mathbf{t}_{he1}(\mathbf{T}_{E2}^{H}\mathbf{B}^{-1}\mathbf{T}_{E1}\boldsymbol{\theta}+\mathbf{T}_{E2}^{H}\mathbf{B}^{-1}\mathbf{h}_{E1})\\
&+\mathbf{t}_{he1}^{H}\mathbf{B}^{-1}\mathbf{t}_{he2}(\mathbf{T}_{E1}^{H}\mathbf{B}^{-1}\mathbf{T}_{E2}\boldsymbol{\theta}+\mathbf{T}_{E1}^{H}\mathbf{B}^{-1}\mathbf{h}_{E2}).\nonumber
\end{align}
After obtaining $\nabla_{\boldsymbol{\theta}}$, we will renew the value $\boldsymbol{\theta}^{(t)}$ of $\boldsymbol{\theta}$ by $\boldsymbol{\theta}^{(t-1)}+\alpha\nabla_{\boldsymbol{\theta}}$ with $\alpha$ being the searching step, which can be obtained by a backtracking line search\cite{conopt}. The detailed process of GA algorithm proposed is listed in Algorithm \ref{algorithm gano}. Thus we can obtain the IRS phase-shift matrix $\boldsymbol{\Theta}$ with $\boldsymbol{\Theta}= \text{diag}\{\boldsymbol{\theta}\}$ .

\begin{algorithm}
\caption{GA algorithm to compute the phase-shift vector $\boldsymbol{\theta}$ using the Max-SR rule}\label{algorithm gano}
\begin{algorithmic}[1]
\STATE Initialize $\boldsymbol{\theta}^{(0)}$, initialize $\mathbf{v}_1$, $\mathbf{v}_2$ based on (\ref{RSV1}) and (\ref{RSV2}), compute $R_s^{(0)}$.
\STATE Set $t=1$, threshold value $\epsilon$.
\REPEAT
\STATE Compute $\nabla_{\boldsymbol{\theta}}^{(t-1)}$ according to (\ref{deltathetano}). Obtain the step size $\alpha^{(t)}$ by backtracking line search.
\STATE $\boldsymbol{\theta}^{(t)}=\boldsymbol{\theta}^{(t-1)}+\alpha^{(t)}\nabla_{\boldsymbol{\theta}}^{(t-1)}$,
reform $\boldsymbol{\theta}^{(t)}=\exp\{j\angle(\boldsymbol{\theta}^{(t)})\}$.
\STATE Compute $R_s^{(t)}$ using $\mathbf{v}_1$, $\mathbf{v}_2$ and $\boldsymbol{\theta}^{(t)}$.
\STATE $t=t+1$.
\UNTIL {$R_s^{(t)}-R_s^{(t-1)}>\epsilon$}
\STATE $\boldsymbol{\theta}^{(t)}$ is the optimal phase-shift vector.
\end{algorithmic}
\end{algorithm}

\subsection{Overall Algorithm}
\begin{algorithm}
\caption{Proposed GAI algorithm}\label{algorithm ais}
\begin{algorithmic}[1]
\STATE Initialize $\mathbf{v}_1^{(0)}$, $\mathbf{v}_2^{(0)}$ and $\boldsymbol{\Theta}^{(0)}$, compute $R_s^{(0)}$ according to (\ref{RSNO}).
\STATE Set $p=0$, threshold $\epsilon$.
\REPEAT
\STATE Given $(\boldsymbol{\Theta}^{(p)},\mathbf{v}_2^{(p)})$, solve problem (\ref{RSV1}) to determine $\mathbf{v}_1^{(p+1)}$ based on the Rayleigh-Ritz theorem.
\STATE Given $(\boldsymbol{\Theta}^{(p)},\mathbf{v}_1^{(p+1)})$, solve problem (\ref{RSV2}) to determine $\mathbf{v}_2^{(p+1)}$ based on the Rayleigh-Ritz theorem.
\STATE Given $(\mathbf{v}_{1}^{(p+1)},\mathbf{v}_2^{(p+1)})$, solve problem (\ref{RST}) to determine $\boldsymbol{\Theta}^{(p+1)}$ based on GA method in Algorithm \ref{algorithm gano}.
\STATE Compute $R_s^{(p+1)}$ using $\mathbf{v}_1^{(p+1)}$, $\mathbf{v}_2^{(p+1)}$ and $\boldsymbol{\Theta}^{(p+1)}$.
\STATE $p=p+1$;
\UNTIL {$R_s^{(p)}-R_s^{(p-1)}\leq\epsilon$}
\STATE $\boldsymbol{\Theta}^{(p)}$, $\mathbf{v}_1^{(p)}$ and $\mathbf{v}_2^{(p)}$ are the optimal value that we need, and $R_s^{(p)}$ is the optimal achievable secrecy rate.
\end{algorithmic}
\end{algorithm}

So far, we have completed the design of beamforming vectors and IRS phase-shift matrix. Our iterative idea can be described as follows:
given  a fixed matrix $\boldsymbol{\Theta}$,  the corresponding beamforming vectors can be computed in a closed-form expression iteratively;
for two given beamforming vectors $\mathbf{v}_1$ and $\mathbf{v}_2$,  the GA method is used  to find the value of IRS phase-shift matrix $\boldsymbol{\Theta}$.
The alternative iteration process among $\mathbf{v}_1$, $\mathbf{v}_2$, and $\boldsymbol{\Theta}$ is repeated until the stop criterion is satisfied, that is, $R_s^{p+1}-R_s^p\leq\epsilon$ with $p$ being the iteration index.
The proposed method is summarized in Algorithm \ref{algorithm ais}.

The computational complexity of Algorithm \ref{algorithm ais} is
\begin{align}\label{GAI-Complexity}
\mathcal{O} \Big(D\big(8N^3+2N+D_1(12M^3K+10M^3+12M^2K+16\nonumber\\
M^2K^2-18M^2+12MK^2+28MK-16M)\log_2{(1/\kappa)}\big)\bigg)
\end{align}
float-point operations~(FLOPs),
where $D$ denotes the maximum number of alternating iterations for Algorithm \ref{algorithm ais}, $D_1$ denotes the maximum iterative number of Algorithm \ref{algorithm gano}, $\kappa$ denotes the accuracy or, in other words, the convergence threshold of backtracking line search, and $\log_2{(1/\kappa)}$ denotes the maximum iterative number of backtracking line search.

\section{Proposed low-complexity NSP-based Max-SR method}\label{S4}

In the previous section, the proposed GAI is general, its computational complexity is still very high because of GA algorithm with lots of FLOPs for obtaining the gradient and stepsize. In this section, we will propose one low-complexity algorithm named NSP to reduce the complexity of the proposed GAI, especially for the case of a large number of IRS elements. In this section, the three beamforming vectors for two CMs and AN are designed well such that any one of them is confined to the NSs of the remaining two channels. This guarantee that two CMs will be not allowed to leak to Eve at the transmitter end, and AN is only transmitted to Eve for interference.

Applying the NSP principle in \cite{HU2016-RDM}, the beamforming vectors $\mathbf{v}_1$ and $\mathbf{v}_2$  can be determined by
\begin{align}\label{nsp}
&\mathbf{H}_{AB}^{H}\mathbf{v}_{1}=\mathbf{0}_{K\times1}, \mathbf{H}_{AE}^{H}\mathbf{v}_{1}=\mathbf{0}_{K\times1},\\
&\mathbf{H}_{AI}\mathbf{v}_{2}=\mathbf{0}_{M\times1}, \mathbf{H}_{AE}^{H}\mathbf{v}_{2}=\mathbf{0}_{K\times1},
\end{align}
which means that $x_1$ is only reflected to users by IRS, and $x_2$ reaches users through the direct path.
The achievable rates from Alice to Bob and to Eve can be expressed as
\begin{align}\label{RBNSP}
R_B=\log_2\left|\mathbf{I}_K+\frac{1}{\sigma^2}\left[\beta_1 P_{s}g_{AIB}\mathbf{H}_{IB}^{H}\boldsymbol{\Theta}\mathbf{H}_{AI}\mathbf{v}_{1}\mathbf{v}_{1}^{H}\left(\mathbf{H}_{IB}^{H}\boldsymbol{\Theta}\mathbf{H}_{AI}\right)^{H}+
\beta_{2}P_{s}g_{AB}\mathbf{H}_{AB}^{H}\mathbf{v}_{2}\mathbf{v}_{2}^{H}\mathbf{H}_{AB}\right]\right|
\end{align}
and
\begin{align}\label{RENSP}
R_E=\log_2\left|\mathbf{I}_K+\frac{\beta_1 P_{s}g_{AIE}\mathbf{H}_{IE}^{H}\boldsymbol{\Theta}\mathbf{H}_{AI}\mathbf{v}_{1}\mathbf{v}_{1}^{H}\left(\mathbf{H}_{IE}^{H}\boldsymbol{\Theta}\mathbf{H}_{AI}\right)^{H}}
{(1-\beta_1-\beta_2)P_s g_{AE}\mathbf{H}_{AE}^{H}\mathbf{P}_{AN}\mathbf{P}_{AN}^{H}\mathbf{H}_{AE}+\sigma^2\mathbf{I}_{K}}\right|.
\end{align}

Let us define two new large channel matrices
\begin{equation}\label{H1}
\mathbf{H}_1 = \left[
\mathbf{H}_{AB}^\ast~
\mathbf{H}_{AE}^\ast
 \right]^T
\end{equation}
and
\begin{equation}\label{H2}
\mathbf{H}_2 = \left[
\mathbf{H}_{AI}^T~
\mathbf{H}_{AE}^\ast
 \right]^T,
\end{equation}
then (\ref{nsp}) can be expressed as
\begin{equation}\label{HV1}
\mathbf{H}_1 \mathbf{v}_1=\mathbf{0}, \mathbf{H}_2 \mathbf{v}_2=\mathbf{0},
\end{equation}
which means the beamforming vectors $\mathbf{v}_1$ and $\mathbf{v}_2$ can be solved by using the {ZF} scheme as $\mathbf{P}_1$ and $\mathbf{P}_2$ are the corresponding projection matrix, where
\begin{align}\label{P1}
\mathbf{P}_1=\mathbf{I}_{N}-\mathbf{H}_{1}^{H}\left[\mathbf{H}_{1}\mathbf{H}_{1}^{H}\right]^{\dagger}\mathbf{H}_{1}
\end{align}
and
\begin{align}\label{P2}
\mathbf{P}_2=\mathbf{I}_{N}-\mathbf{H}_{2}^{H}\left[\mathbf{H}_{2}\mathbf{H}_{2}^{H}\right]^{\dagger}\mathbf{H}_{2}.
\end{align}

For convenience of derivation below, let us define two new vectors $\mathbf{w}_1\in \mathbb{C}^{N\times 1}$ and $\mathbf{w}_2\in \mathbb{C}^{N\times 1}$
\begin{align}\label{vtt}
\mathbf{v}_1=\mathbf{P}_1\mathbf{w}_1, \mathbf{v}_2=\mathbf{P}_2\mathbf{w}_2.
\end{align}
As for the condition (\ref{P02}) of the problem (\ref{P0}),  we rewrite it by applying (\ref{vtt}), that is,
$\mathbf{w}_1^{H}\mathbf{P}_1^{H}\mathbf{P}_1\mathbf{w}_1=1$ and $\mathbf{w}_2^{H}\mathbf{P}_2^{H}\mathbf{P}_2\mathbf{w}_2=1$.   (\ref{yban}) and (\ref{yean}) are rewritten as follows
\begin{align}
\mathbf{y}_B
&=\sqrt{\beta_1 P_sg_{AIB}}\mathbf{H}^{H}_{IB}\boldsymbol{\Theta}\mathbf{H}_{AI}\mathbf{v}_1x_{1} \label{ybnsp}+\sqrt{\beta_2 P_sg_{AB}}\mathbf{H}^{H}_{AB}\mathbf{v}_2x_{2}+\mathbf{n}_{B}\\
&=\sqrt{\beta_1 P_sg_{AIB}}\mathbf{H}^{H}_{IB}\boldsymbol{\Theta}\mathbf{H}_{AI}\mathbf{P}_1\mathbf{w}_1x_{1} +\sqrt{\beta_2 P_sg_{AB}}\mathbf{H}^{H}_{AB}\mathbf{P}_2\mathbf{w}_2x_{2}+\mathbf{n}_{B},\nonumber
\end{align}
and
\begin{align}
\mathbf{y}_E
&=\sqrt{\beta_1P_s{g_{AIE}}}
\mathbf{H}^{H}_{IE}\boldsymbol{\Theta}\mathbf{H}_{AI}\mathbf{v}_{1}x_{1}\label{yensp} +\sqrt{(1-\beta_1-\beta_2)P_s{g_{AE}}}\mathbf{H}^{H}_{AE}\mathbf{P}_{AN}\mathbf{z}+\mathbf{n}_{E}\\
&=\sqrt{\beta_1P_s{g_{AIE}}}
\mathbf{H}^{H}_{IE}\boldsymbol{\Theta}\mathbf{H}_{AI}\mathbf{P}_1\mathbf{w}_1 x_{1}+\sqrt{(1-\beta_1-\beta_2)P_s{g_{AE}}}\mathbf{H}^{H}_{AE}\mathbf{P}_{AN}\mathbf{z}+\mathbf{n}_{E}.\nonumber
\end{align}
In what follows, we can calculate the beamforming vectors and IRS phase-shift matrix by calculating $\mathbf{w}_{1}$, $\mathbf{w}_2$ and $\boldsymbol{\Theta}$ alternatively.

\subsection{Optimization of beamforming vectors  given IRS phase-shift matrix $\boldsymbol{\Theta}$}
Substituting   (\ref{vtt}) in (\ref{RBNSP})  yields
\begin{align}\label{RBNSPV}
R_B=\log_2|\mathbf{I}_{K}+\mathbf{A}_1\mathbf{w}_1\mathbf{w}_1^H\mathbf{A}_1^{H}+\mathbf{A}_2\mathbf{w}_2\mathbf{w}_2^H\mathbf{A}_2^{H}|,
\end{align}
where $\mathbf{A}_1=\frac{\sqrt{\beta_1 P_s g_{AIB}}}{\sigma}\mathbf{H}_{IB}^{H}\boldsymbol{\Theta}\mathbf{H}_{AI}\mathbf{P}_1$, $\mathbf{A}_2=\frac{\sqrt{\beta_2 P_s g_{AB}}}{\sigma}\mathbf{H}_{AB}^{H}\mathbf{P}_2$,
Similarlly, substituting   (\ref{vtt}) in  (\ref{RENSP}) yields
\begin{align}\label{RENSPV}
R_E&=\log_2|\mathbf{I}_{K}+\mathbf{A}_3\mathbf{w}_1\mathbf{w}_1^H\mathbf{A}_3^{H}\mathbf{B}^{-1}|\overset{(b)}{=}\log_2\left(1+\mathbf{w}_1^H\mathbf{A}_3^{H}\mathbf{B}^{-1}\mathbf{A}_3\mathbf{w}_1 \right),
\end{align}
where $\mathbf{A}_3=\frac{\sqrt{\beta_1 P_s g_{AIE}}}{\sigma}\mathbf{H}_{IE}^{H}\boldsymbol{\Theta}\mathbf{H}_{AI}\mathbf{P}_1$, $\mathbf{B}$ owns the same definition as (\ref{B}), and $(b)$ holds due to $|\mathbf{I}_M+\mathbf{X}\mathbf{Y}|=|\mathbf{I}_N+\mathbf{Y}\mathbf{X}|$ for $\mathbf{X}\in \mathbb{C}^{M\times N}$ and $\mathbf{Y}\in \mathbb{C}^{N\times M}$.
Then the NSP-based Max-SR can be formulated as follows:
\begin{subequations}\label{P1NSP}
\begin{align}
\mathrm{(P1):}&\max_{\mathbf{w}_1,\mathbf{w}_2,\boldsymbol{\Theta}}~~R_s(\mathbf{w}_1,\mathbf{w}_2,\boldsymbol{\Theta})=(\ref{RBNSPV})-(\ref{RENSPV})\label{P1NSP0}\\
&~~~~\text{s.t.}~~\mathbf{w}_1^{H}\mathbf{P}_1^{H}\mathbf{P}_1\mathbf{w}_1=1,~\mathbf{w}_2^{H}\mathbf{P}_2^{H}\mathbf{P}_2\mathbf{w}_2=1,\label{P1NSP1}\\
&~~~~~~~~~~(\ref{P02}).\label{P1NSP2}
\end{align}
\end{subequations}
It is clear to see that $R_B$ in (\ref{RBNSPV}) is related to $\mathbf{w}_1$, $\mathbf{w}_2$ and $\boldsymbol{\Theta}$, while $R_E$ in (\ref{RENSPV}) is only related to $\mathbf{w}_2$ and $\boldsymbol{\Theta}$.
Since the expression of (\ref{RBNSPV}) is similar to (\ref{RBNO}), (\ref{RBNSPV}) can be expressed as the function of $\mathbf{w}_1$ in (\ref{RBNSPW1}) with known $\mathbf{w}_2$ and $\boldsymbol{\Theta}$, and the function of $\mathbf{w}_2$ in (\ref{RBNSPW2}) with known $\mathbf{w}_1$ and $\boldsymbol{\Theta}$
\begin{align}
{R}_B(\mathbf{w}_1)=\log_2|\mathbf{I}_{K}+\mathbf{A}_2\mathbf{w}_2\mathbf{w}_2^H\mathbf{A}_2^{H}|+\label{RBNSPW1}\log_2\left(1+\mathbf{w}_1^H\mathbf{A}_1^{H}(\mathbf{I}_{K}+\mathbf{A}_2\mathbf{w}_2\mathbf{w}_2^H\mathbf{A}_2^{H})^{-1}\mathbf{A}_1\mathbf{w}_1\right),
\end{align}
and
\begin{align}
{R}_B(\mathbf{w}_2)=\log_2|\mathbf{I}_{K}+\mathbf{A}_1\mathbf{w}_1\mathbf{w}_1^H\mathbf{A}_1^{H}|+\label{RBNSPW2}\log_2\left(1+\mathbf{w}_2^H\mathbf{A}_2^{H}(\mathbf{I}_{K}+\mathbf{A}_1\mathbf{w}_1\mathbf{w}_1^H\mathbf{A}_1^{H})^{-1}\mathbf{A}_2\mathbf{w}_2\right),
\end{align}
respectively.
Since  ${R}_B(\mathbf{w}_1)$ in  (\ref{RBNSPW1}) is independent of $\mathbf{w}_2$, the NSP-based Max-SR of optimizing $\mathbf{w}_1$  is casted as
\begin{subequations}\label{PNSPV1}
\begin{align}
\mathrm{(P1-1):}&\max_{\mathbf{w}_1}~~\frac{\mathbf{w}_1^{H}\tilde{\mathbf{A}}_{1}\mathbf{w}_1}{\mathbf{w}_1^{H}\tilde{\mathbf{B}}_{1}\mathbf{w}_1} \label{PNSPV11}\\
&~\text{s.t.}~~\mathbf{w}_1^{H}\mathbf{P}_1^{H}\mathbf{P}_1\mathbf{w}_1=1,\label{PNSPV12}
\end{align}
\end{subequations}
where
\begin{align}
\tilde{\mathbf{A}}_{1}=\mathbf{P}_1^{H}\mathbf{P}_1+\mathbf{A}_{1}^{H}(\mathbf{I}_{K}+\mathbf{A}_{2}\mathbf{w}_2\mathbf{w}_2^{H}\mathbf{A}_{2}^{H})^{-1}\mathbf{A}_{1}
\end{align}
and
\begin{align}
\tilde{\mathbf{B}}_{1}=\mathbf{P}_1^{H}\mathbf{P}_1+\mathbf{A}_{3}^{H}\mathbf{B}^{-1}\mathbf{A}_{3}.
\end{align}
Since $\frac{\mathbf{w}_1^{H}\tilde{\mathbf{A}}_{1}\mathbf{w}_1}{\mathbf{w}_1^{H}\tilde{\mathbf{B}}_{1}\mathbf{w}_1} $ is insensitive to the scaling of $\mathbf{w}_1$,  via ignoring the constraint on $\mathbf{w}_1$, we will find a general solution, and then scale it to satisfy
\begin{align}\label{stwppw1}
\mathbf{w}_1^{H}\mathbf{P}_1^{H}\mathbf{P}_1\mathbf{w}_1\leq1.
\end{align}
It can be observed that the optimization problem in (\ref{PNSPV1}) belongs to the type of nonlinear fractional optimization problem. To solve this problem, we introduce the Dinkelbach method, and then transform it into a DC programming similar to \cite{dinkel}. Since the numerator and denominator of the objective function in problem (\ref{PNSPV1}) are convex, we introduce $\nu$ into it and transform it as
\begin{align}\label{dinwAw}
\mathbf{w}_1^{H}\tilde{\mathbf{A}}_{1}\mathbf{w}_1-\nu{\mathbf{w}_1^{H}\tilde{\mathbf{B}}_{1}\mathbf{w}_1}.
\end{align}
Then (\ref{PNSPV11}) can be achieved if and only if
\begin{align}\label{PNSPV11nu}
\max_{\mathbf{w}_1\in\mathbb{D}}&~~~\mathbf{w}_1^{H}\tilde{\mathbf{A}}_{1}\mathbf{w}_1-\nu^\ast{\mathbf{w}_1^{H}\tilde{\mathbf{B}}_{1}\mathbf{w}_1}=\mathbf{w}_1^{*H}\tilde{\mathbf{A}}_{1}\mathbf{w}_1^*-\nu^\ast{\mathbf{w}_1^{*H}\tilde{\mathbf{B}}_{1}\mathbf{w}_1^*}=0,
\end{align}
for $\mathbf{w}_1^{H}\tilde{\mathbf{A}}_{1}\mathbf{w}_1\geq0$ and ${\mathbf{w}_1^{H}\tilde{\mathbf{B}}_{1}\mathbf{w}_1}\geq0$, $\forall\mathbf{w}_1\in\mathbb{D}$, where $\mathbb{D}$ denotes the feasible domain of the problem (\ref{PNSPV1}). This transformation can be proved in \cite{dinkel}.
Therefore, we can rewrite the optimization problem (\ref{PNSPV1}) as
\begin{align}\label{PNSPV1t}
\mathrm{(P1-1.1):}\max_{\mathbf{w}_1,\nu}~~~\mathbf{w}_1^{H}\tilde{\mathbf{A}}_{1}\mathbf{w}_1-\nu{\mathbf{w}_1^{H}\tilde{\mathbf{B}}_{1}\mathbf{w}_1}~~~~~~\text{s.t.}~~~~(\ref{stwppw1}).
\end{align}

However, problem (\ref{PNSPV1t}) is still not convex in terms of $\mathbf{w}_1$ due to the fact that its objective function is the difference of two convex functions, which is nonconvex. Hence, we linearize  the objective function $\mathbf{w}_1^{H}\tilde{\mathbf{A}}_{1}\mathbf{w}_1$ by  the first term of its Taylor series expansion at a given vector of $\tilde{\mathbf{w}}_1$ as follows\cite{liqxax}
\begin{align}\label{xAx}
\mathbf{w}_1^{H}\tilde{\mathbf{A}}_{1}\mathbf{w}_1\geq2\Re\{\tilde{\mathbf{w}}_1^{H}\tilde{\mathbf{A}}_{1}\mathbf{w}_1\}-\tilde{\mathbf{w}}_1^{H}\tilde{\mathbf{A}}_{1}\tilde{\mathbf{w}}_1.
\end{align}
Then the problem (\ref{PNSPV1}) can be rewritten as
\begin{align}\label{PNSPV1x}
\mathrm{(P1-1.2):}&\max_{\mathbf{w}_1,\nu}~2\Re\{\tilde{\mathbf{w}}_1^{H}\tilde{\mathbf{A}}_{1}\mathbf{w}_1\}-\tilde{\mathbf{w}}_1^{H}\tilde{\mathbf{A}}_{1}\tilde{\mathbf{w}}_1-\nu{\mathbf{w}_1^{H}\tilde{\mathbf{B}}_{1}\mathbf{w}_1}~~~~~~~~~\text{s.t.}~~(\ref{stwppw1}).
\end{align}
which is  a convex optimization problem. Then it can be readily solved by \cite{conopt}.

The  optimization subproblem of NSP-based Max-SR with respect to $\mathbf{w}_2$   can be modeled as
\begin{align}\label{PNSPV2}
\mathrm{(P1-2):}\max_{\mathbf{w}_2}~~\mathbf{w}_2^{H}\tilde{\mathbf{A}}_{2}\mathbf{w}_2 ~~~~~~~~\text{s.t.}~~\mathbf{w}_2^{H}\mathbf{P}_2^{H}\mathbf{P}_2\mathbf{w}_2=1,
\end{align}
where
\begin{align}
\tilde{\mathbf{A}}_{2}=\mathbf{P}_2^{H}\mathbf{P}_2+\mathbf{A}_{2}^{H}(\mathbf{I}_{K}+\mathbf{A}_{1}\mathbf{w}_1\mathbf{w}_1^{H}\mathbf{A}_{1}^{H})^{-1}\mathbf{A}_{2}.
\end{align}
Similarly, the constraint can be scaled as (\ref{stwppw1}).
Since $\tilde{\mathbf{A}}_{2}\succeq\mathbf{0}$, $\mathbf{w}_2^{H}\tilde{\mathbf{A}}_{2}\mathbf{w}_2$ is a convex function with respect to $\mathbf{w}_2$, we can get the following inequality by performing the first-order Taylor expansion on $\mathbf{w}_2^{H}\tilde{\mathbf{A}}_{2}\mathbf{w}_2$ at the point $\tilde{\mathbf{w}}_2$ like (\ref{xAx}).
Then the problem (\ref{PNSPV2}) can be rewritten as
\begin{align}\label{PNSPV2x}
\mathrm{(P1-2.1):}\max_{\mathbf{w}_2}~~2\Re\{\tilde{\mathbf{w}}_2^{H}\tilde{\mathbf{A}}_{2}\mathbf{w}_2\}-\tilde{\mathbf{w}}_2^{H}\tilde{\mathbf{A}}_{2}\tilde{\mathbf{w}}_2~~~~~~~\text{s.t.}~~\mathbf{w}_2^{H}\mathbf{P}_2^{H}\mathbf{P}_2\mathbf{w}_2\leq1.
\end{align}
We can see that the objective function in the optimization problem (\ref{PNSPV2x}) is concave and the constraint is convex. Thus (\ref{PNSPV2x}) is a convex optimization problem, which can be solved by \cite{conopt}.
\subsection{Optimization of IRS phase-shift matrix $\boldsymbol{\Theta}$ with given beamforming vectors}
Now, we optimize the IRS phase-shift matrix $\boldsymbol{\Theta}$ by using NSP-based Max-SR method.  By applying ($\ref{thetade}$),  (\ref{RBNSP}) and (\ref{RENSP}) are represented as
\begin{align}
R_B(\boldsymbol{\theta})=\log_2|\mathbf{I}_{K}+\mathbf{T}_{B1}\boldsymbol{\theta}\boldsymbol{\theta}^{H}\mathbf{T}_{B1}^{H}+\mathbf{h}_{B2}\mathbf{h}_{B2}^{H}|,\label{RBNSPT}
\end{align}
and
\begin{align}
R_E(\boldsymbol{\theta})=\log_2|\mathbf{I}_{K}+\mathbf{T}_{E1}\boldsymbol{\theta}\boldsymbol{\theta}^{H}\mathbf{T}_{E1}^{H}\mathbf{B}^{-1}|,\label{RENSPT}
\end{align}
where $\mathbf{T}_{B1}$, $\mathbf{h}_{B2}$, $\mathbf{T}_{E1}$, and $\mathbf{B}$ have the same forms as (\ref{TB1}), (\ref{hb}), (\ref{TE1}), and (\ref{B}).
Then the subproblem to optimize $\boldsymbol{\Theta}$ can be equivalently changed as to optimize the IRS phase-shift vector $\boldsymbol{\theta}$, formulated as,
\begin{align}\label{PNSPT}
\mathrm{(P1-3):}\max_{\boldsymbol{\theta}}~~(\ref{RBNSPT})-(\ref{RENSPT})~~~~~~~~~\text{s.t.}~~~ (\ref{thetacon}).
\end{align}
Due to the fact that $|\mathbf{X}\mathbf{Y}|=|\mathbf{X}||\mathbf{Y}|$ and $|\mathbf{I}_M+\mathbf{X}\mathbf{Y}|=|\mathbf{I}_N+\mathbf{Y}\mathbf{X}|$ for $\mathbf{X}\in \mathbb{C}^{M\times N}$ and $\mathbf{Y}\in \mathbb{C}^{N\times M}$, (\ref{RBNSPT}) and (\ref{RENSPT}) can be rewritten as
\begin{align}
R_B(\boldsymbol{\theta})&=\log_2(1+\boldsymbol{\theta}^{H}\mathbf{T}_{B1}^{H}(\mathbf{I}_{K}+\mathbf{h}_{B2}\mathbf{h}_{B2}^{H})^{-1}\mathbf{T}_{B1}\boldsymbol{\theta})\label{RBNSPT1}+\log_2|\mathbf{I}_{K}+\mathbf{h}_{B2}\mathbf{h}_{B2}^{H}|,
\end{align}
and
\begin{align}
R_E(\boldsymbol{\theta})&=\log_2(1+\boldsymbol{\theta}^{H}\mathbf{T}_{E1}^{H}\mathbf{B}^{-1}\mathbf{T}_{E1}\boldsymbol{\theta}).\label{RENSPT1}
\end{align}
Since $\log_2|\mathbf{I}_K+\mathbf{h}_{B2}\mathbf{h}_{B2}^{H}|$ is independent of $\boldsymbol{\theta}$, problem (\ref{PNSPT}) can be formulated as
\begin{align}\label{PRSNSPT}
\mathrm{(P1-3.1):}&\max_{\boldsymbol{\theta}}~~\frac{\boldsymbol{\theta}^{H}\tilde{\mathbf{T}}_B\boldsymbol{\theta}}
{\boldsymbol{\theta}^{H}\tilde{\mathbf{B}}_E\boldsymbol{\theta}}~~~~~~~~~~\text{s.t.}~~~(\ref{thetacon}),
\end{align}
where
\begin{align}
&\tilde{\mathbf{T}}_B=\frac{1}{M}\mathbf{I}_{M}+\mathbf{T}_{B1}^{H}(\mathbf{I}_{K}+\mathbf{h}_{B2}\mathbf{h}_{B2}^{H})^{-1}\mathbf{T}_{B1},\label{TBtlide}\\
&\tilde{\mathbf{B}}_E=\frac{1}{M}\mathbf{I}_{M}+\mathbf{T}_{E1}^{H}\mathbf{B}^{-1}\mathbf{T}_{E1}.\label{BEtlide}
\end{align}

Rewrite problem  (\ref{PRSNSPT}) as
\begin{align}\label{PRSd1}
\mathrm{(P1-3.2):}\min_{\boldsymbol{\theta}}~~\frac{\boldsymbol{\theta}^{H}\tilde{\mathbf{B}}_E\boldsymbol{\theta}}
{\boldsymbol{\theta}^{H}\tilde{\mathbf{T}}_B\boldsymbol{\theta}}~~~~~~~~~\text{s.t.}~~~(\ref{thetacon}).
\end{align}
Obviously, the above optimization problem belongs to fractional programming. Introducing a new parameter $\mu\geqslant0$ forms the corresponding parametric program as follows:
\begin{align}\label{PRSd2}
\mathrm{(P1-3.3):}\min_{\boldsymbol{\theta}}~~\boldsymbol{\theta}^{H}\tilde{\mathbf{B}}_E\boldsymbol{\theta}-\mu
\boldsymbol{\theta}^{H}\tilde{\mathbf{T}}_B\boldsymbol{\theta}~~~~~~~~~~~~\text{s.t.}~~~(\ref{thetacon}).
\end{align}
As \cite{dinkel} showed, the optimal solution to problem (\ref{PRSd2}) is the unique root of $\boldsymbol{\theta}^{H}\tilde{\mathbf{B}}_E\boldsymbol{\theta}-\mu
\boldsymbol{\theta}^{H}\tilde{\mathbf{T}}_B\boldsymbol{\theta}=0$.
Without the constant mode constraint of $\boldsymbol{\theta}$, this kind of problem can be solved by SDR as problem (\ref{PNSPV1x}) performs.
In this case, we minimize an upper bound of its objective function following \cite{SJ-wAw} as
\begin{align}\label{upperbound}
&\boldsymbol{\theta}^{H}\tilde{\mathbf{B}}_E\boldsymbol{\theta}-\mu\boldsymbol{\theta}^{H}\tilde{\mathbf{T}}_B\boldsymbol{\theta}
=\boldsymbol{\theta}^{H}(\tilde{\mathbf{B}}_E-\mu\tilde{\mathbf{T}}_B)\boldsymbol{\theta}\\
&\leqslant\lambda_{\max}(\boldsymbol{\Psi})\|\boldsymbol{\theta}\|^2-2\Re\{\boldsymbol{\theta}^{H}
\big(\lambda_{\max}(\boldsymbol{\Psi})\mathbf{I}_M-\boldsymbol{\Psi}\big)\tilde{\boldsymbol{\theta}}\}+\tilde{\boldsymbol{\theta}}^{H}\big(\lambda_{\max}(\boldsymbol{\Psi})\mathbf{I}_M-\boldsymbol{\Psi}\big)\tilde{\boldsymbol{\theta}},\nonumber
\end{align}
where $\boldsymbol{\Psi}=\tilde{\mathbf{B}}_E-\mu\tilde{\mathbf{T}}_B$, $\tilde{\boldsymbol{\theta}}$ is the solution to $\boldsymbol{\theta}$ obtained in the previous iteration of the alternating algorithm. Since $|\boldsymbol{\theta}_i|^2=1$ and $\|\boldsymbol{\theta}\|^2=M$, $\lambda_{\max}(\boldsymbol{\Psi})\|\boldsymbol{\theta}\|^2$ and $\tilde{\boldsymbol{\theta}}^{H}\big(\lambda_{\max}(\boldsymbol{\Psi})\mathbf{I}_M-\boldsymbol{\Psi}\big)\tilde{\boldsymbol{\theta}}$ are determined here.
Then the simplified optimization problem reduces to
\begin{align}\label{PRSd3}
\mathrm{(P1-3.4):}\max_{\boldsymbol{\theta}}~~\Re\{\boldsymbol{\theta}^{H}\boldsymbol{\delta}\}~~~~~~~~~\text{s.t.}~~~(\ref{thetacon}),
\end{align}
where $\boldsymbol{\delta}=\big(\lambda_{\max}(\boldsymbol{\Psi})\mathbf{I}_M-\boldsymbol{\Psi}\big)\tilde{\boldsymbol{\theta}}$. In this case, $\Re\{\boldsymbol{\theta}^{H}\boldsymbol{\delta}\}$ is maximized when the phases of $\theta_i$ and $\delta_i$ are equal, where $\delta_i$ is the $i$-th element of $\boldsymbol{\delta}$.
Thus the optimal solution to the problem with given $\mu$ is
\begin{align}\label{thetaast}
\boldsymbol{\theta}^\ast(\mu)=[e^{j\arg(\delta_1)},\cdots,e^{j\arg(\delta_M)}]^{T}.
\end{align}
Substituting $\boldsymbol{\theta}^\ast(\mu)$ into the objective function of problem (\ref{PRSd2}), we have the result  $\varphi^{\ast}(\mu)$. Since $\varphi^{\ast}(\mu)$ is a strictly decreasing function for the optimal $\boldsymbol{\theta}$, with $\varphi^{\ast}(0)>0$ and $\varphi^{\ast}(+\infty)<0$, which has been confirmed in \cite{SH-SRMIRS}, the optimal $\mu^{\ast}$ can be found by $\varphi^{\ast}(\mu^\ast)=0$ via bisection search. Thus we can obtain the solution to $\boldsymbol{\theta}$ by $\boldsymbol{\theta}^\ast(\mu^{\ast})$.
The above problem has a closed form, which is more convenient for implementation and requires much lower complexity especially for large $M$.


\subsection{Overall Algorithm}
The proposed NSP algorithm is divided  into two parts: the  beamforming vectors and the IRS phase-shift matrix. The iterative idea can be described as follows:
for given matrix $\boldsymbol{\Theta}$, anyone of the beamforming vectors can be expressed as an unknown vector  multiplied by a known matrix, which can be computed by CVX iteratively as the other is fixed;
for given two beamforming vectors $\mathbf{v}_1$ and $\mathbf{v}_2$, the closed-form expression of IRS phase-shift vector $\boldsymbol{\theta}$ can be expressed as (\ref{thetaast}).
The alternative iterations among $\mathbf{v}_1$, $\mathbf{v}_2$ and $\boldsymbol{\Theta}$ is repeated until the stop criterion satisfies, that is, $R_s^{p+1}-R_s^p\leq\epsilon$ with $p$ being the iteration index.
The proposed method is summarized in Algorithm \ref{algorithm nsp}.

{ The computational complexity of Algorithm \ref{algorithm nsp} is
\begin{align}\label{NSP-Complexity}
&\mathcal{O}\Big(L\big( 2\sqrt{2}[{(N+1)}^3 + N^2(N+1)]\text{ln}({1}/{\epsilon})
+L_1 (M^3+4M^2\nonumber\\
&K-2M-2MK+4MK^2+K^2) \log_2({(\lambda_{max}-\lambda_{min})}/{\epsilon})\big)\Big)
\end{align}
FLOPs, where $L$ denotes the maximum number of alternating iterations, $L_1$ denotes the iterative number of the subproblem (P1-3), $\epsilon$ denotes the accuracy or the convergence threshold of the algorithm, and $\lambda_{max}$ and $\lambda_{min}$  are the upper-bound and lower-bound of bisection method, respectively. $\log_2({(\lambda_{max}-\lambda_{min})}/{\epsilon})$ is the maximum iterative number of bisection search.}

  Compared with  the complexity of the proposed GAI in (\ref{GAI-Complexity}) , the complexity of the proposed NSP in (\ref{NSP-Complexity}) is greatly reduced especially for large $M$ by taking the convergence analysis in Section V into account. This is the benefit of NSP. However, the NSP is only suitable for the case that three streams are transmitted separately and directively, and requires that the number of transmit antennas is greater than the number of receive antennas. This is its limit.
Additionally, compared to the GAI , the proposed NSP algorithm will suffer from a performance loss due to its strict NS constraints. This will reduce the spatial multiplexing gain of CMs.  In summary, the proposed NSP can strike an appreciated good balance between SR performance and computational complexity.

\begin{algorithm}
\caption{Proposed NSP algorithm}\label{algorithm nsp}
\begin{algorithmic}[1]
\STATE Initialize $\mathbf{v}_1^{(0)}$, $\mathbf{v}_2^{(0)}$ and $\boldsymbol{\Theta}^{(0)}$, compute $R_s^{(0)}$ according to (\ref{RBNSP}) and (\ref{RENSP}).
\STATE Set $p=0$, threshold $\epsilon$.
\REPEAT
\STATE Given $(\boldsymbol{\Theta}^{(p)},\mathbf{v}_2^{(p)})$ and (\ref{vtt}), solve problem (\ref{PNSPV1x}) to determine $\mathbf{v}_1^{(p+1)}$.
\STATE Given $(\boldsymbol{\Theta}^{(p)},\mathbf{v}_1^{(p+1)})$ and (\ref{vtt}), solve problem (\ref{PNSPV2x}) to determine $\mathbf{v}_2^{(p+1)}$.
\STATE Given $(\mathbf{v}_{1}^{(p+1)},\mathbf{v}_2^{(p+1)})$ and (\ref{vtt}), $\boldsymbol{\theta}^{(p+1)}$ can be determined by (\ref{thetaast}), $\boldsymbol{\Theta}^{(p+1)} = \text{diag}\{\boldsymbol{\theta}^{(p+1)}\}$.
\STATE Compute $R_s^{(p+1)}$ using $\mathbf{v}_1^{(p+1)}$, $\mathbf{v}_2^{(p+1)}$ and $\boldsymbol{\Theta}^{(p+1)}$.
\STATE $p=p+1$;
\UNTIL {$R_s^{(p)}-R_s^{(p-1)}\leq\epsilon$}
\STATE $\boldsymbol{\Theta}^{(p)}$, $\mathbf{v}_1^{(p)}$ and $\mathbf{v}_2^{(p)}$ are the optimal value that we need, and $R_s^{(p)}$ is the optimal achievable secrecy rate.
\end{algorithmic}
\end{algorithm}

\section{Simulation and Discussion}\label{S5}
In this section, we provide numeral results to examine the performance of our proposed algorithms.
As for the MIMO system model, the array response is modeled as $\mathbf{a}_{t}(\theta_t)\in\mathbb{C}^{n_t\times1}$, with $[\mathbf{a}_{t}(\theta_t)]_{n_{ti}}=\exp(-j2\pi(n_{ti}-1)d_{A}\cos\theta_{ti}/\lambda)$, where $\theta_t\in[0,~\pi)$ denotes the angle-of-arrival (AoA), and $\mathbf{a}_{r}(\theta_r)\in\mathbb{C}^{n_r\times1}$, with $[\mathbf{a}_{r}(\theta_r)]_{n_{ri}}=\exp(-j2\pi(n_{ri}-1)d_{A}\cos\theta_{ri}/\lambda)$, where $\theta_r\in[0,~\pi)$ denotes the angle-of-departure (AoD). Both transmit array at Alice and receive array at Bob are uniformly spaced linear arrays with element pacing $d_A=\lambda/2$. The LoS channel matrix can be expressed as $\mathbf{H}=\mathbf{a}_{r}(\theta_r)\mathbf{a}_{t}^{H}(\theta_t)$.
The path loss model is given by $g_{TR}=\big(\frac{c}{4\pi d_{TR}f}\big)^2$, where $d_{TR}$ denotes the distance between the transmitter and the receiver. Under this model, the path loss coefficient $g_{AB},~g_{AE},~g_{AIB}$ and $g_{AIE}$ can be derived respectively.

Simulation parameters are set as follows : $P_s=30$ dBm, $\sigma_B^2=\sigma_E^2=\sigma^2=-40$ dBm.
 $N=16$, $K=4$. The distances of Alice-to-IRS link, Alice-to-Bob link, and Alice-to-Eve link are set as $d_{AI}=10$ m, $d_{AB}=100$ m and $d_{AE}=50$ m, respectively. The AoDs of each channel are set as $\theta^{t}_{AI}=\pi/6$, $\theta^{t}_{AB}=11\pi/36$ and $\theta^{t}_{AE}=\pi/3$, respectively. With given AoDs and distances of each channel, the AoAs and distances of IRS-to-Bob link and IRS-to-Eve link can be determined, thus the channel matrix can be derived respectively. The PA factors are set as $\beta_1 = \beta_2 = 0.4,~ \beta_3 = 0.2$. As for the algorithm setup, the convergence thresholds in terms of the relative increment in the objective value are set as tolerance of $\epsilon=10^{-4}$.


\subsection{Convergence Behaviour of Proposed Algorithms}
\begin{figure}
  \centering
  \includegraphics[width=0.5\textwidth]{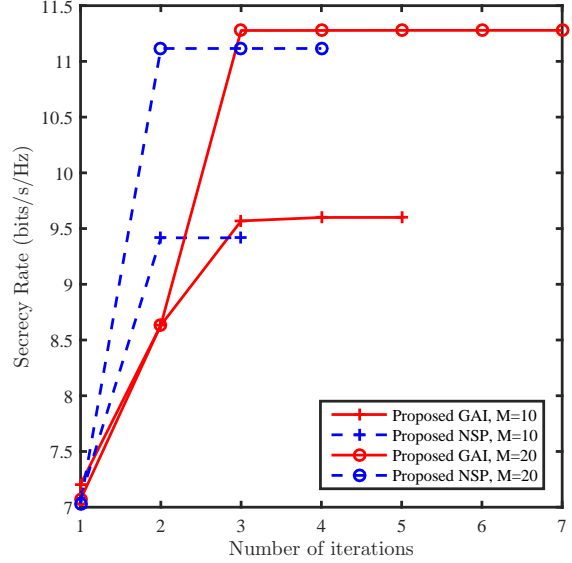}\\
  \caption{Convergence of proposed algorithm at different number of IRS phase-shift elements. }\label{SR_CON}
\end{figure}

First, by simulation, we make an investigation of the convergence behaviour of the proposed GAI in Algorithm \ref{algorithm ais} and NSP in Algorithm \ref{algorithm nsp}. Fig.~\ref{SR_CON} shows the SR versus the number of iterations for various number of phase shifter.,~i.e.,~for $M=10, 20$. It can be seen from the figure that GAI  requires about 4 iterations to converge the SR ceil, while the proposed NSP  requires about 3 iterations to converge. Thus, we make a conclusion that the proposed NSP has a more rapid convergence rate than GAI. Using the convergence results in Fig.~\ref{SR_CON},  the complexity (\ref{GAI-Complexity}) of GAI and complexity (\ref{GAI-Complexity}) of NSP reduce to the magnitude orders $40M^3$ and $3M^3$ FLOPs respectively as $M$ goes to large-scale. Clearly, the complexity of NSP is far lower than that of GAI.

\subsection{Performance Comparison}
In this subsection, we compare our proposed algorithms to the following benchmark schemes:
\begin{enumerate}
 \item \textbf{No-IRS}: Obtain the maximum SR by optimizing the beamforming vectors with the IRS phase-shift matrix set to zero,~i.e.,~$\boldsymbol{\Theta}=\boldsymbol{0}_{M \times M}$.

 \item \textbf{Random Phase}: Obtain the maximum SR by optimizing the beamforming vectors with all the phase for each reflection element uniformly and independently generated from $[0,2\pi)$.


 \item \textbf{IRS with Single CBS}: Obtain the maximum SR by Algorithm \ref{algorithm ais} with single CBS, as $\beta_1 = 0, \beta_2 = 1-\beta_3$ or $\beta_2 = 0, \beta_1 = 1-\beta_2$. In this case, we also fix PA factor of the AN as $\beta_3 = 0.2$.
\end{enumerate}

\subsubsection{Impact of the Number of IRS Phase-shift Elements}
\begin{figure}
  \centering
  \includegraphics[width=0.5\textwidth]{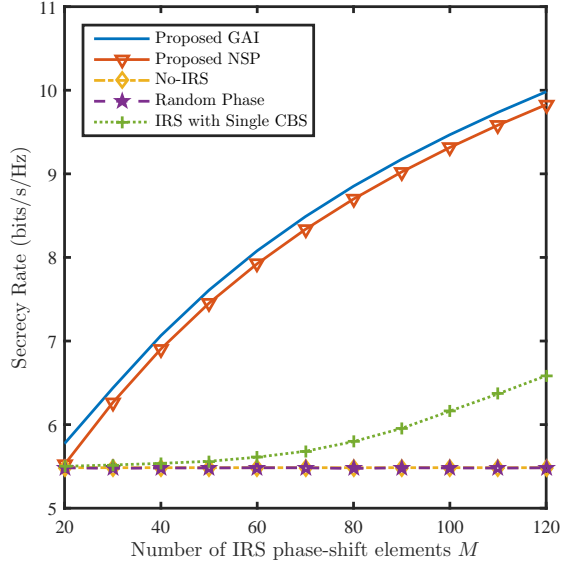}\\
  \caption{Secrecy rate versus the number of IRS phase-shift elements $M$ in the low-SNR regime ( $d_{AB}=300$~m). }\label{RS_M_lowSNR}
\end{figure}
\begin{figure}
  \centering
  \includegraphics[width=0.5\textwidth]{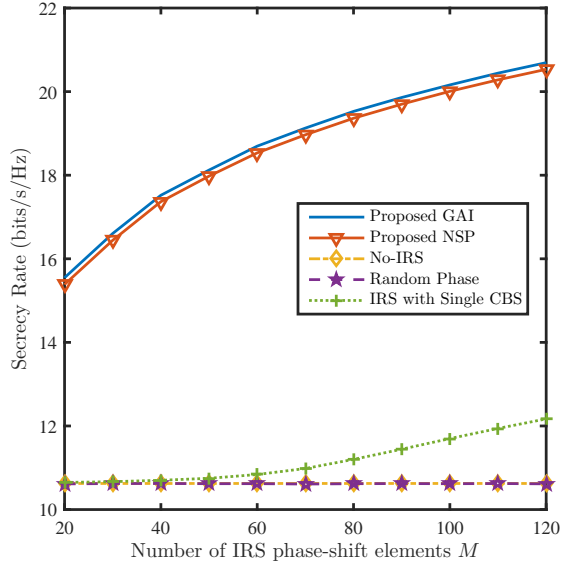}\\
  \caption{Secrecy rate versus the number of IRS phase-shift elements $M$ in the high-SNR regime ( $d_{AB}=50$~m). }\label{RS_M_highSNR}
\end{figure}

For comparison, we consider two scenarios of Alice-to-Bob distance given by $d_{AB}=300$ m and $d_{AB}=50$ m, which correspond to the \emph{low-SNR regime} and \emph{high-SNR regime}, respectively. For these two cases, the SR performance versus the number of reflecting elements $M$ for the proposed algorithms and the benchmark schemes are presented as Fig.~\ref{RS_M_lowSNR} and Fig.~\ref{RS_M_highSNR}.

From Fig.~\ref{RS_M_lowSNR} and Fig.~\ref{RS_M_highSNR}, it can be seen that the proposed  two schemes GAI and NSP can improve the SR performance whether in the \emph{low-SNR regime} or the \emph{high-SNR regime}.  As the number of IRS elements increases,  the SR gains achieved by GAI and NSP over no-IRS, random phase and IRS with single CBS grow gradually and become more significant.

Compared with the No-IRS scheme and Random-Phase scheme, the IRS phase-shift-optimization schemes (i.e.,~GAI,~NSP) performs much better, especially with a large value of $M$. This reveals the importance of the optimization of the phase-shift design. Even with a  value of $M$ as $M=30$, our proposed scheme can also perform better than that scheme without the IRS phase-shift-optimization (e.g., by 17.3\% in the \emph{low-SNR regime} and 56.3\% in the \emph{high-SNR regime}).

Under the condition that the total power is equally allocated  between two independent CBSs, the proposed GAI  performs a bit better than the proposed NSP. This shows that the proposed  NSP  scheme sacrifices a
 little SR performance by an obvious computational complexity reduction.

Compared with the case of IRS with single CBS, the SR performance in the case of dual CM stream plus AN (i.e.,~GAI and NSP scheme) is much better whether in the \emph{low-SNR regime} or in the \emph{high-SNR regime} (e.g.,~by 16.6\% higher in the \emph{low-SNR regime} and 55.6\% higher in the \emph{high-SNR regime} when $M=30$). This proves the superiority of our proposed schemes in the dual CM stream case due to the diversity gain in LOS channel.
Furthermore, even with IRS aided, Alice transmitting single CBS can not achieve better security performance than the case  without IRS, unless the IRS equips with more phase-shift elements. This is because the path loss of the IRS-forward link is more serious than the direct link in LOS channel. If there is no more IRS phase-shift elements, IRS may not forward single CBS to the legitimate user more strongly. In this case, it is suggested to transmit dual CBSs or more CM streams with IRS aided, which requires more in-depth researches in the future.

On the other hand, the performance gap between our proposed schemes and other schemes increases as the IRS phase-shift elements $M$ and receive SNR increases, which reveals the superiority of our proposed schemes.

\subsubsection{Impact of the IRS Location}
\begin{figure}
  \centering
  \includegraphics[width=0.5\textwidth]{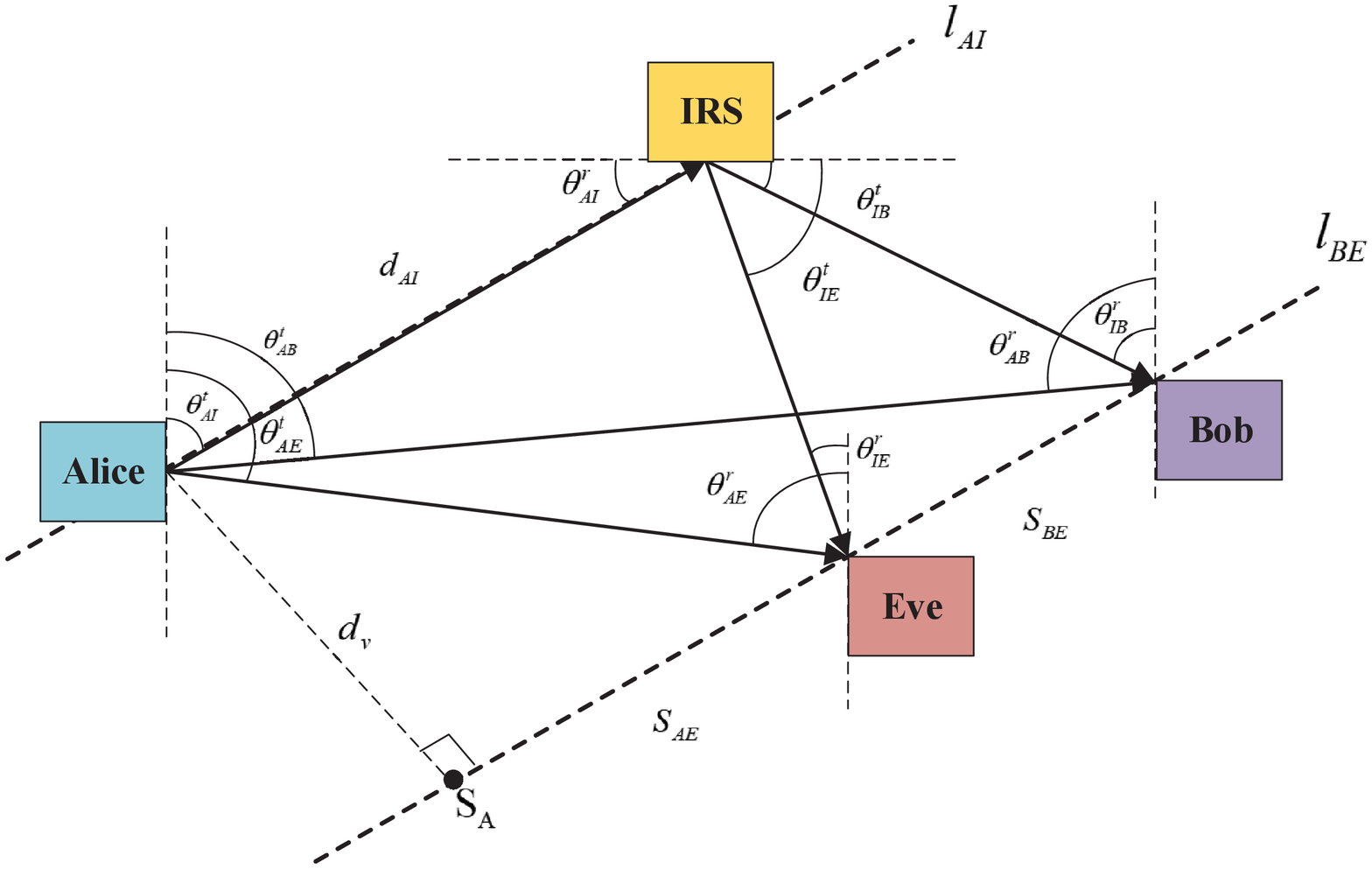}\\
  \caption{Special scenario for adjusting the position of IRS. ($l_{AI}$ is parallel to $l_{BE}$) }\label{systemmodel_ai}
\end{figure}

With fixed positions of Alice, Bob and Eve, the IRS position only depends on the AoD $\theta_{AI}^{t}$ and the distance $d_{AI}$ of Alice-to-IRS link. To simplify the analysis, assume that Alice and IRS are on a straight line $l_{AI}$ parallel to the straight line $l_{BE}$ with Bob and Eve. The distances and AoDs of Alice-to-Bob link and Alice-to-Eve link are computed as before, thus $\theta_{AI}^{t}$ can be determined as (\ref{thetaait}).
\begin{align}\label{thetaait}
\begin{cases}
\theta_{AI}^{t} = \theta_{AB}^{t} - \arcsin\left( \frac{d_{AE}}{d_{BE}}\sin\theta_{BAE} \right),\\
d_{BE} = \sqrt{ d_{AB}^{2}+d_{AE}^{2}-2d_{AB}d_{AE}\cos\theta_{BAE}},\\
\theta_{BAE} = \theta_{AE}^{t} - \theta_{AB}^{t}.
\end{cases}
\end{align}
The vertical distance $d_v$ of the two lines $l_{AI}$ and $l_{BE}$ can be computed as $d_v = d_{AE} \sin\left(\theta_{AE}^{t} - \theta_{AI}^{t}\right)$. Fig.~\ref{systemmodel_ai} shows the location scenario.
Define the point $\text{S}_{A}$ is the projection point on $l_{BE}$, which means $l_{\text{A}\text{S}_{A}}\perp l_{BE}$. Then the distances between $\text{S}_{A}$ and Eve, $\text{S}_{A}$ and Bob can be expressed as $d_{\text{S}_{A}\text{E}}=\sqrt{d_{AE}^2 - d_v^2}$,  $d_{\text{S}_{A}\text{B}}=\sqrt{d_{AB}^2 - d_v^2}$, respectively. Based on the above conditions, $\theta_{AI}^{t} = 5\pi/18$, the distance of $d_{\text{S}_{A}\text{E}}$ and $d_{\text{S}_{A}\text{B}}$ can be calculated as $d_{\text{S}_{A}\text{E}} = 49.2$~m, $d_{\text{S}_{A}\text{B}} = 99.6$~m.

\begin{figure}
  \centering
  \includegraphics[width=0.5\textwidth]{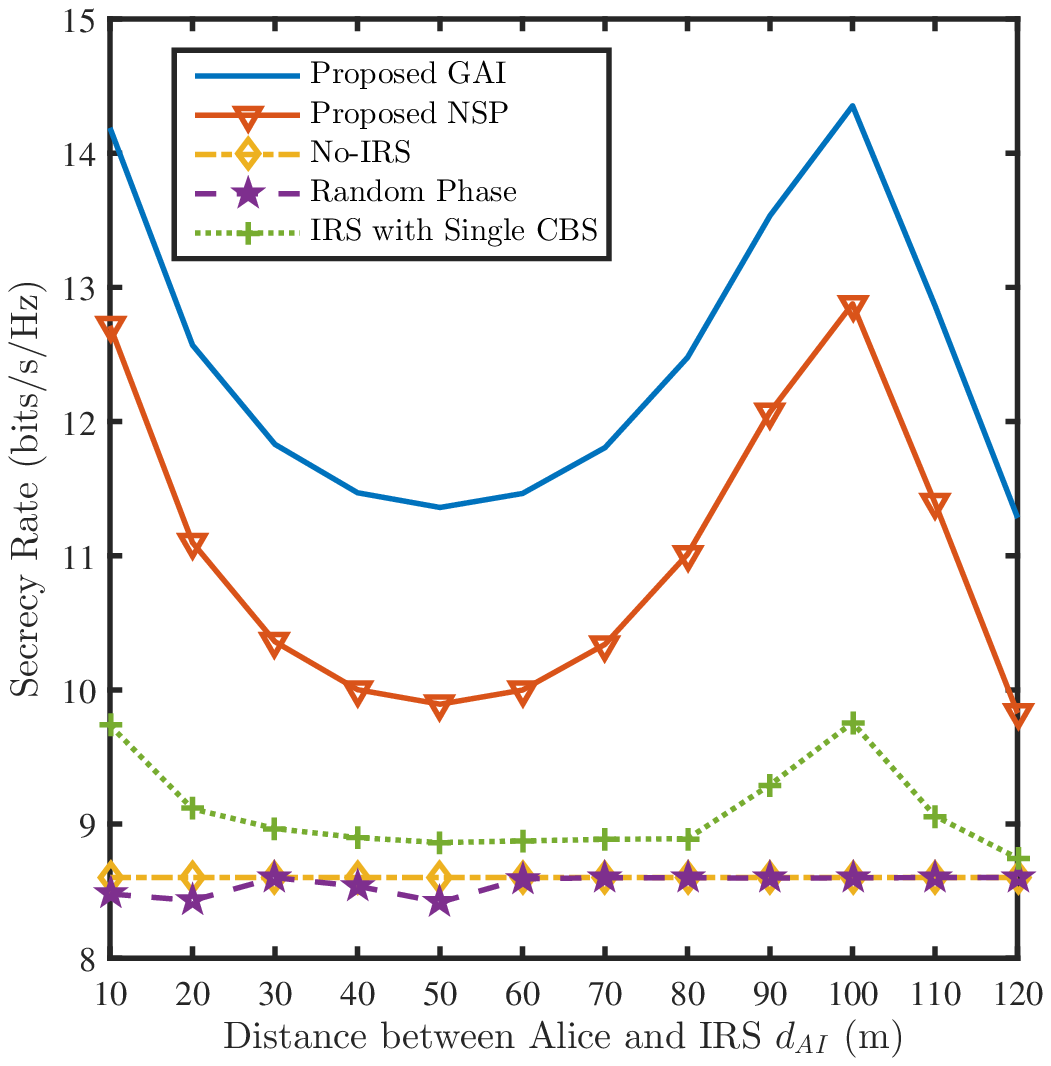}
  \caption{Secrecy rate versus distance between Alice and IRS $d_{AI}$ ($M=80$). }\label{RS_d_ai}
\end{figure}

Fig.~\ref{RS_d_ai} depicts the SR versus $d_{AI}$ when $M=80$ as shown in the scenario in (\ref{systemmodel_ai}). Here, IRS moves from the position of Alice along the line  $l_{AI}$ near Bob.
 As IRS gets closer to Eve but still far away from Bob, the achievable SR decreases gradually.  When IRS is on top of Eve, the minimum SR value is available. In this moment, when IRS is the nearest to Eve, Eve has the strongest eavesdropping ability.  As IRS moves away along the line $l_{AI}$ from Eve, it get closer and closer to Bob, the SR value increases up to the largest until IRS is on top of Bob. Furthermore, as IRS moves away along the line $l_{AI}$ Bob, both Eve and Bob get less energy reflected from IRS, thus the SR decreases gradually.

\section{Conclusion}\label{S6}
In this paper, we have made an extensive investigation of secure transmit beamforming and phase shifting at IRS in a secure IRS-based DM Networks, where  two parallel independent CBSs are transmitted from Alice to Bob with multiple receive antennas.  Using the criterion of Max-SR,  two alternating iterative algorithms, GAI and NSP, have been proposed. The former is of high-performance and the latter is of low-complexity. From simulation, we find the IRS can make a dramatic enhancement on the SR of DM by using two CBSs compared to single CBS. For example, with the aid of IRS,  the proposed two methods can approximately double the SR of existing method with single CBS in the case of medium-scale and large-scale IRS.  Additionally, the impact of IRS position on SR is also analyzed in the simulation. It is recommended that the IRS is placed close to the transmitter or the target receiver to achieve  a higher SR performance. Moreover, the optimal position of IRS also exists.

\ifCLASSOPTIONcaptionsoff
  \newpage
\fi

\bibliographystyle{IEEEtran}
\bibliography{IEEEfull,cite}

\begin{thebibliography}{10}
\providecommand{\url}[1]{#1}
\csname url@samestyle\endcsname
\providecommand{\newblock}{\relax}
\providecommand{\bibinfo}[2]{#2}
\providecommand{\BIBentrySTDinterwordspacing}{\spaceskip=0pt\relax}
\providecommand{\BIBentryALTinterwordstretchfactor}{4}
\providecommand{\BIBentryALTinterwordspacing}{\spaceskip=\fontdimen2\font plus
\BIBentryALTinterwordstretchfactor\fontdimen3\font minus
  \fontdimen4\font\relax}
\providecommand{\BIBforeignlanguage}[2]{{%
\expandafter\ifx\csname l@#1\endcsname\relax
\typeout{** WARNING: IEEEtran.bst: No hyphenation pattern has been}%
\typeout{** loaded for the language `#1'. Using the pattern for}%
\typeout{** the default language instead.}%
\else
\language=\csname l@#1\endcsname
\fi
#2}}
\providecommand{\BIBdecl}{\relax}
\BIBdecl

\bibitem{massivemimo}
E.~G. {Larsson}, O.~{Edfors}, F.~{Tufvesson}, and T.~L. {Marzetta}, ``Massive
  {MIMO} for next generation wireless systems,'' \emph{IEEE Commun. Mag.},
  vol.~52, no.~2, pp. 186--195, Feb. 2014.

\bibitem{TS-MMWAVE}
T.~S. {Rappaport}, S.~{Sun}, R.~{Mayzus}, H.~{Zhao}, Y.~{Azar}, K.~{Wang},
  G.~N. {Wong}, J.~K. {Schulz}, M.~{Samimi}, and F.~{Gutierrez}, ``Millimeter
  wave mobile communications for {5G} cellular: It will work!'' \emph{IEEE
  Access}, vol.~1, pp. 335--349, 2013.

\bibitem{hbsurvey}
I.~{Ahmed}, H.~{Khammari}, A.~{Shahid}, A.~{Musa}, K.~S. {Kim}, E.~{De
  Poorter}, and I.~{Moerman}, ``A survey on hybrid beamforming techniques in
  {5G}: Architecture and system model perspectives,'' \emph{IEEE Commun.
  Surveys Tuts.}, vol.~20, no.~4, pp. 3060--3097, Fourthquarter 2018.

\bibitem{WANG-OFDMA1}
H.~{Zhu} and J.~{Wang}, ``Chunk-based resource allocation in {OFDMA} systems -
  part {I}: chunk allocation,'' \emph{IEEE Trans. Commun.}, vol.~57, no.~9, pp.
  2734--2744, Sep. 2009.

\bibitem{WANG-OFDMA2}
H.~{Zhu} and J.~{Wang}, ``Chunk-based resource allocation in {OFDMA} systems -
  part {II}: joint chunk, power and bit allocation,'' \emph{IEEE Trans.
  Commun.}, vol.~60, no.~2, pp. 499--509, Feb. 2012.

\bibitem{liqxax}
Q.~{Li}, Q.~{Zhang}, and J.~{Qin}, ``Secure relay beamforming for {SWIPT} in
  amplify-and-forward two-way relay networks,'' \emph{IEEE Trans. Veh.
  Technol.}, vol.~65, no.~11, pp. 9006--9019, Nov. 2016.

\bibitem{WuQ-viewGreen5G}
Q.~{Wu}, G.~Y. {Li}, W.~{Chen}, D.~W.~K. {Ng}, and R.~{Schober}, ``An overview
  of sustainable green {5G} networks,'' \emph{IEEE Wireless Commun.}, vol.~24,
  no.~4, pp. 72--80, Aug. 2017.

\bibitem{MMGreen5G}
M.~M. {Mowla}, I.~{Ahmad}, D.~{Habibi}, and Q.~V. {Phung}, ``A green
  communication model for {5G} systems,'' \emph{IEEE Trans. Green Commun.
  Netw.}, vol.~1, no.~3, pp. 264--280, Sep. 2017.

\bibitem{Wyner1975}
A.~D. Wyner, ``The wire-tap channel,'' \emph{Bell. Syst. Tech. J.}, vol.~54,
  no.~8, pp. 1355--1387, Oct. 1975.

\bibitem{AN-wiretap-YSH}
S.~{Yan}, N.~{Yang}, I.~{Land}, R.~{Malaney}, and J.~{Yuan}, ``Three
  artificial-noise-aided secure transmission schemes in wiretap channels,''
  \emph{IEEE Trans. Veh. Technol.}, vol.~67, no.~4, pp. 3669--3673, Apr. 2018.

\bibitem{DM-PA}
M.~P. {Daly} and J.~T. {Bernhard}, ``Directional modulation technique for
  phased arrays,'' \emph{IEEE Trans. Antennas Propag.}, vol.~57, no.~9, pp.
  2633--2640, Sep. 2009.

\bibitem{YD-avector}
Y.~{Ding} and V.~F. {Fusco}, ``A vector approach for the analysis and synthesis
  of directional modulation transmitters,'' \emph{IEEE Trans. Antennas
  Propag.}, vol.~62, no.~1, pp. 361--370, Jan. 2014.

\bibitem{sun2020energy-efficient}
L.~{Sun}, J.~{Li}, Y.~{Zhang}, Y.~{Wang}, L.~{Gui}, F.~{Li}, H.~{Li},
  Z.~{Zhuang}, and F.~{Shu}, ``Energy-efficient alternating iterative secure
  structure of maximizing secrecy rate for directional modulation networks,''
  \emph{Physical Commun.}, vol.~38, p. 100949, 2020.

\bibitem{LJY-DM}
J.~{Li}, L.~{Xu}, P.~{Lu}, T.~{Liu}, Z.~{Zhuang}, J.~{Hu}, F.~{Shu}, and
  J.~{Wang}, ``Performance analysis of directional modulation with
  finite-quantized rf phase shifters in analog beamforming structure,''
  \emph{IEEE Access}, vol.~7, pp. 97\,457--97\,465, 2019.

\bibitem{HU2016-RDM}
J.~{Hu}, F.~{Shu}, and J.~{Li}, ``Robust synthesis method for secure
  directional modulation with imperfect direction angle,'' \emph{IEEE Commun.
  Lett.}, vol.~20, no.~6, pp. 1084--1087, Jun. 2016.

\bibitem{WU2016-RDMBC}
F.~{Shu}, X.~{Wu}, J.~{Li}, R.~{Chen}, and B.~{Vucetic}, ``Robust synthesis
  scheme for secure multi-beam directional modulation in broadcasting
  systems,'' \emph{IEEE Access}, vol.~4, pp. 6614--6623, Oct. 2016.

\bibitem{MU-MIMO-ZHU}
F.~{Shu}, W.~{Zhu}, X.~{Zhou}, J.~{Li}, and J.~{Lu}, ``Robust secure
  transmission of using main-lobe-integration-based leakage beamforming in
  directional modulation {MU-MIMO} systems,'' \emph{IEEE Syst. J.}, vol.~12,
  no.~4, pp. 3775--3785, Dec. 2018.

\bibitem{DM-XU}
F.~{Shu}, L.~{Xu}, J.~{Wang}, W.~{Zhu}, and Z.~{Xiaobo},
  ``Artificial-noise-aided secure multicast precoding for directional
  modulation systems,'' \emph{IEEE Trans. Veh. Technol.}, vol.~67, no.~7, pp.
  6658--6662, Jul. 2018.

\bibitem{DOA-QIN}
F.~{Shu}, Y.~{Qin}, T.~{Liu}, L.~{Gui}, Y.~{Zhang}, J.~{Li}, and Z.~{Han},
  ``Low-complexity and high-resolution {DOA} estimation for hybrid analog and
  digital massive {MIMO} receive array,'' \emph{IEEE Trans. Commun.}, vol.~66,
  no.~6, pp. 2487--2501, Jun. 2018.

\bibitem{RFDA-HJS}
J.~{Hu}, S.~{Yan}, F.~{Shu}, J.~{Wang}, J.~{Li}, and Y.~{Zhang},
  ``Artificial-noise-aided secure transmission with directional modulation
  based on random frequency diverse arrays,'' \emph{IEEE Access}, vol.~5, pp.
  1658--1667, 2017.

\bibitem{Wu-SPTDM}
F.~{Shu}, X.~{Wu}, J.~{Hu}, J.~{Li}, R.~{Chen}, and J.~{Wang}, ``Secure and
  precise wireless transmission for random-subcarrier-selection-based
  directional modulation transmit antenna array,'' \emph{IEEE J. Sel. Areas
  Commun.}, vol.~36, no.~4, pp. 890--904, Apr. 2018.

\bibitem{wuqq-zongshu}
Q.~{Wu} and R.~{Zhang}, ``Towards smart and reconfigurable environment:
  Intelligent reflecting surface aided wireless network,'' \emph{IEEE Commun.
  Mag.}, vol.~58, no.~1, pp. 106--112, 2020.

\bibitem{huangc-2018GC}
C.~{Huang}, G.~C. {Alexandropoulos}, A.~{Zappone}, M.~{Debbah}, and C.~{Yuen},
  ``Energy efficient multi-user {MISO} communication using low resolution large
  intelligent surfaces,'' in \emph{2018 IEEE Globecom Workshops (GC Wkshps)},
  2018, pp. 1--6.

\bibitem{huangc-2019}
C.~{Huang}, A.~{Zappone}, G.~C. {Alexandropoulos}, M.~{Debbah}, and C.~{Yuen},
  ``Reconfigurable intelligent surfaces for energy efficiency in wireless
  communication,'' \emph{IEEE Trans. Wireless Commun.}, vol.~18, no.~8, pp.
  4157--4170, 2019.

\bibitem{wuqq1}
Q.~{Wu} and R.~{Zhang}, ``Intelligent reflecting surface enhanced wireless
  network via joint active and passive beamforming,'' \emph{IEEE Trans. on
  Wireless Commun.}, vol.~18, no.~11, pp. 5394--5409, 2019.

\bibitem{wuqq3-discrete}
Q.~{Wu} and R.~{Zhang}, ``Beamforming optimization for wireless network aided
  by intelligent reflecting surface with discrete phase shifts,'' \emph{IEEE
  Trans. Commun.}, vol.~68, no.~3, pp. 1838--1851, 2020.

\bibitem{IRS-ZY-MIMO}
S.~{Zhang} and R.~{Zhang}, ``Capacity characterization for intelligent
  reflecting surface aided {MIMO} communication,'' \emph{arXiv preprint}, vol.
  arXiv:1910.01573, 2019.

\bibitem{IRS-mmMIMO}
V.~{Jamali}, A.~M. {Tulino}, G.~{Fischer}, R.~{Muller}, and R.~{Schober},
  ``Intelligent reflecting and transmitting surface aided millimeter wave
  massive {MIMO},'' \emph{arXiv preprint}, vol. arXiv:1902.07670, 2019.

\bibitem{IRS-ZY-MIMOtw}
Y.~{Zhang}, C.~{Zhong}, Z.~{Zhang}, and W.~{Lu}, ``Sum rate optimization for
  two way communications with intelligent reflecting surface,'' \emph{IEEE
  Commun. Lett.}, vol.~24, no.~5, pp. 1090--1094, 2020.

\bibitem{IRS-CR-XG}
X.~{Guan}, Q.~{Wu}, and R.~{Zhang}, ``Joint power control and passive
  beamforming in {IRS}-assisted spectrum sharing,'' \emph{IEEE Commun. Lett.},
  vol.~24, no.~7, pp. 1553--1557, 2020.

\bibitem{IRS-CR-LZ}
L.~{Zhang}, Y.~{Wang}, W.~{Tao}, Z.~{Jia}, T.~{Song}, and C.~{Pan},
  ``Intelligent reflecting surface aided {MIMO} cognitive radio systems,''
  \emph{IEEE Trans. Veh. Technol.}, Online.

\bibitem{IRS-SL-UAV}
S.~{Li}, B.~{Duo}, X.~{Yuan}, Y.~{Liang}, and M.~{Di Renzo}, ``Reconfigurable
  intelligent surface assisted {UAV} communication: Joint trajectory design and
  passive beamforming,'' \emph{IEEE Wireless Commun. Lett.}, vol.~9, no.~5, pp.
  716--720, 2020.

\bibitem{SHI2019SWIPT}
W.~{Shi}, X.~{Zhou}, L.~{Jia}, Y.~{Wu}, F.~{Shu}, and J.~{Wang}, ``Enhanced
  secure wireless information and power transfer via intelligent reflecting
  surface,'' \emph{arXiv preprint}, vol. arXiv:1911.01001, 2019.

\bibitem{IRS-MIMO-SWIPT}
C.~{Pan}, H.~{Ren}, K.~{Wang}, M.~{Elkashlan}, A.~{Nallanathan}, J.~{Wang}, and
  L.~{Hanzo}, ``Intelligent reflecting surface aided {MIMO} broadcasting for
  simultaneous wireless information and power transfer,'' \emph{IEEE J. Sel.
  Areas Commun.}, Online.

\bibitem{IRS-O-number}
E.~{Bjornson}, O.~{Ozdogan}, and E.~G. {Larsson}, ``Intelligent reflecting
  surface versus decode-and-forward: How large surfaces are needed to beat
  relaying?'' \emph{IEEE Wireless Commun. Lett.}, vol.~9, no.~2, pp. 244--248,
  2020.

\bibitem{IRS-pathloss-DN}
W.~{Tang}, M.~{Chen}, X.~{Chen}, J.~{Dai}, Y.~{Han}, M.~D. {Renzo}, Y.~{Zeng},
  S.~{Jin}, Q.~{Cheng}, and T.~J. {Cui}, ``Wireless communications with
  reconfigurable intelligent surface: Path loss modeling and experimental
  measurement,'' \emph{arXiv preprint}, vol. arXiv:1911.0326, 2019.

\bibitem{IRS-O-pathloss}
O.~{Ozdogan}, E.~{Bjornson}, and E.~G. {Larsson}, ``Intelligent reflecting
  surfaces: Physics, propagation, and pathloss modeling,'' \emph{IEEE Wireless
  Commun. Lett.}, vol.~9, no.~5, pp. 581--585, 2020.

\bibitem{IRS-CM-secure}
M.~{Cui}, G.~{Zhang}, and R.~{Zhang}, ``Secure wireless communication via
  intelligent reflecting surface,'' \emph{IEEE Wireless Commun. Lett.}, vol.~8,
  no.~5, pp. 1410--1414, 2019.

\bibitem{SH-SRMIRS}
H.~{Shen}, W.~{Xu}, S.~{Gong}, Z.~{He}, and C.~{Zhao}, ``Secrecy rate
  maximization for intelligent reflecting surface assisted multi-antenna
  communications,'' \emph{IEEE Commun. Lett.}, vol.~23, no.~9, pp. 1488--1492,
  Sep. 2019.

\bibitem{IRS-LD-secureMIMO}
L.~{Dong} and H.~{Wang}, ``Secure {MIMO} transmission via intelligent
  reflecting surface,'' \emph{IEEE Wireless Commun. Lett.}, vol.~9, no.~6, pp.
  787--790, 2020.

\bibitem{IRS-LD-ANsecureMIMO}
L.~{Dong} and H.~{Wang}, ``Enhancing secure {MIMO} transmission via intelligent
  reflecting surface,'' \emph{IEEE Trans. Wireless Commun.}, Online.

\bibitem{IRS-AN}
X.~{Guan}, Q.~{Wu}, and R.~{Zhang}, ``Intelligent reflecting surface assisted
  secrecy communication: Is artificial noise helpful or not?'' \emph{IEEE
  Wireless Commun. Lett.}, vol.~9, no.~6, pp. 778--782, 2020.

\bibitem{IRS-JSAC-ROBUSTAN}
X.~{Yu}, D.~{Xu}, Y.~{Sun}, D.~W.~K. {Ng}, and R.~{Schober}, ``Robust and
  secure wireless communications via intelligent reflecting surfaces,''
  \emph{IEEE J. Sel. Areas Commun.}, Online.

\bibitem{PAN_SEC_MIMO}
S.~{Hong}, C.~{Pan}, H.~{Ren}, K.~{Wang}, and A.~{Nallanathan},
  ``Artificial-noise-aided secure {MIMO} wireless communications via
  intelligent reflecting surface,'' \emph{arXiv preprint}, vol.
  arXiv:2002.07063, 2020.

\bibitem{conopt}
S.~{Boyd} and L.~{Vandenberghe}, \emph{Convex Optimization}.\hskip 1em plus
  0.5em minus 0.4em\relax Cambridge, U.K.: Cambridge Univ. Press, 2004.

\bibitem{dinkel}
W.~{Dinkelbach}, ``On nonlinear fractional programming,'' \emph{Manage. Sci.},
  vol.~13, no.~7, pp. 492--498, Mar. 1967.

\bibitem{SJ-wAw}
J.~{Song}, P.~{Babu}, and D.~P. {Palomar}, ``Optimization methods for designing
  sequences with low autocorrelation sidelobes,'' \emph{IEEE Trans. Signal
  Process.}, vol.~63, no.~15, pp. 3998--4009, Aug 2015.

\end{thebibliography}

\end{document}